\documentclass[a4paper,fleqn, sort&compress]{cas-sc}

\usepackage[utf8]{inputenc}
\usepackage[english]{babel}
\usepackage[square,numbers]{natbib}

\usepackage{placeins}
\usepackage{amsmath}
\usepackage{graphicx}
\usepackage{tcolorbox}	
\usepackage{relsize}
\usepackage{tabu,multirow}
\usepackage{epsfig}
\usepackage{array}
\usepackage{amssymb}
\usepackage{latexsym}
\usepackage{xcolor}
\usepackage{subcaption}
\usepackage{soul}
\usepackage[export]{adjustbox}
\usepackage{setspace}
\usepackage{caption}
\usepackage{float}
\usepackage{tikz}
\usetikzlibrary{positioning, shapes, arrows}
\usepackage{epsfig}
\usepackage{siunitx}
\usepackage{miller}
\usepackage[noabbrev,capitalise]{cleveref}

\newcommand{\boldface}[1]{\boldsymbol{#1}}  

\newcommand{\bfo}{\boldface{o}}
\newcommand{\bfp}{\boldface{p}}
\newcommand{\bfq}{\boldface{q}}

\newcommand{\bfz}{\boldface{z}}

\newcommand{\bfD}{\boldface{D}}

\newcommand{\bfF}{\boldface{F}}

\newcommand{\bfSigma}{\boldsymbol{\Sigma}}

\newcommand{\calZ}{\mathcal{Z}}
\newcommand{\itbf}[1]{\textit{\textbf{#1}}}
\newcommand{\T}{^{\mathrm{T}}} 

\newlength{\boxwidth}
\def\dd{\;\!\mathrm{d}}

\def\btheorem{\begin{theorem}}
\def\etheorem{\end{theorem}}
\def\blemma{\begin{lemma}}
\def\elemma{\end{lemma}}
\def\bproposition{\begin{proposition}}
\def\eproposition{\end{proposition}}
\def\bcorollary{\begin{corollary}}
\def\ecorollary{\end{corollary}}
\def\bdefinition{\begin{definition}}
\def\edefinition{\end{definition}}
\def\bexample{\begin{example}}
\def\eexample{\end{example}}
\def\bremark{\begin{remark}}
\def\eremark{\end{remark}}

\newcommand{\be}{\begin{equation*}}
\newcommand{\ee}{\end{equation*}}

\newcommand{\beq}{\begin{eqnarray*}}
\newcommand{\eeq}{\end{eqnarray*}}
\newcommand{\bem}{\begin{multline}}
\newcommand{\eem}{\end{multline}}
\newcommand{\ba}{\begin{align*}}
\newcommand{\ea}{\end{align*}}

\newcommand{\fp}[2]{\frac{\partial #1}{\partial #2} }

\DeclareMathOperator{\gammaGBsub}{\gamma_\textsf{gb}^\textsf{sub}}
\DeclareMathOperator{\gammaGBfull}{\gamma_\textsf{gb}^\textsf{full}}
\DeclareMathOperator{\gammaGBlim}{\gamma_\textsf{gb}^\textsf{lim}}
\DeclareMathOperator{\gammaGB}{\gamma_\textsf{gb}}
\DeclareMathOperator{\gammaSurf}{\gamma_\textsf{surf}}
\DeclareSIUnit\angstrom{\text{Å}}

\begin{document}

\shorttitle{Finite temperature grain boundary properties}
\shortauthors{Sp\'inola et~al.}

\title[mode=title]{Finite-temperature grain boundary properties from quasistatic atomistics}
\author[eth]{Miguel Sp\'inola}[orcid=0000-0002-5180-6149]
\author[eth]{Shashank Saxena}[orcid=0000-0002-5242-9103]
\author[iit]{Prateek Gupta}[orcid=0000-0003-3666-0257]
\author[isu]{Brandon Runnels}[orcid=0000-0003-3043-5227]
\cormark[1]
\ead{brunnels@iastate.edu} 
\author[eth]{Dennis M. Kochmann}[orcid=0000-0002-9112-6615]
\cormark[1]
\ead{dmk@ethz.ch}
\cortext[1]{Corresponding authors}
\address[eth]{Mechanics \& Materials Lab, ETH Z\"urich, Switzerland}
\address[isu]{Department of Aerospace Engineering, Iowa State University, Ames, IA USA}
\address[iit]{Department of Applied Mechanics, Indian Institute of Technology, Delhi, India}

\begin{abstract}
Grain boundary (GB) properties greatly influence the mechanical, electrical, and thermal response of polycrystalline materials. Most computational studies of GB properties at finite temperatures use molecular dynamics (MD), which is computationally expensive, limited in the range of accessible timescales, and requires cumbersome techniques like thermodynamic integration to estimate free energies. This restricts the reasonable computation (without incurring excessive computational expense) of GB properties to regimes that are often unrealistic,  such as zero temperature or extremely high strain rates. Consequently, there is a need for simulation methodology that avoids the timescale limitations of MD, while providing reliable estimates of GB properties. The Gaussian Phase-Packet (GPP) method is a temporal coarse-graining technique that can predict relaxed atomic structures at finite temperature in the quasistatic limit. This work applies GPP, combined with the quasiharmonic approximation for computing the free energy, to the problem of determining the free energy and shear coupling factor of grain boundaries in metals over a range of realistic temperatures. Validation is achieved by comparison to thermodynamic integration, which confirms that the presented approach captures relaxed-energy GB structures and shear coupling factors at finite temperature with a high degree of accuracy.
\end{abstract}


\maketitle

\section{Introduction}
Mechanical \cite{laporte2009intermediate}, thermal \cite{dong2014relative} , electrical \cite{bishara2021understanding} or chemical \cite{bechtle2009grain} properties of polycrystalline materials are greatly influenced by grain boundary (GB) properties \cite{balluffi}. For example, the mechanical strength of polycrystals can both increase and decrease due to the presence of grain boundaries \cite{raabe2014grain}. A well-known weakening mechanism is the segregation of alloying solute atoms around GBs, which promotes intergranular fracture, thus embrittling the material. On the other hand, \citet{bechtle2009grain} showed that increasing the number of special boundaries (mostly twin boundaries) in nickel to reduce hydrogen-induced intergranular embrittlement can result in an increased tensile ductility. Another work by \citet{KUZMINA2015182} showed that the driving force for solute segregation is linked to the GB energy. By lowering the GB energy, the driving force for segregation is decreased, delaying the embrittlement temperature. GBs can also act as strengthening agents for a material by impeding dislocation propagation. One of the well-known strengthening mechanisms is the Hall-Petch effect \cite{dao2007toward}, by which the yield strength of a solid can be increased by reducing the grain size. This increases the density of GB area, resulting in an increasing competition between GB phenomena and dislocation-mediated plasticity. For example, \citet{taali2022grain} showed that finer grains in copper sheets increased tensile strength and by tuning grain size across the thickness of the sheet tensile strength and ductility could be controlled. Hence, by engineering the properties of GBs it is possible to substantially enhance mechanical properties. This proves the need for an increased understanding and modeling of GB properties.

Experimental methods are essential to the understanding of GBs, but even modern experimental techniques suffer from significant challenges when exploring GB intrinsic behavior \cite{rollett2017recrystallization} and cannot always provide quantitative results \cite{dao2007toward}. For example, experimental measurements of GB mobility during recrystallization are difficult to perform, and some GB structures have rare natural occurrence or are experimentally inaccessible \cite{rollett2017recrystallization}. Therefore, the theoretical and computational modeling of GBs is of great scientific and technological interest. Modeling of GBs goes back to the classical low-angle model of \citet{read1950dislocation} in the 1950s and the introduction of the coincident site lattice (CSL) and structure identification \cite{kronberg1949secondary}. \citet{friedel1953etude} performed one of the first GB atomistic energy calculations, using an experimentally calibrated interatomic potential for aluminum and iron. Other early studies \cite{weins1970structure, weins1971computer} showed the multiplicity of GB energy metastable states. 
In atomistic simulations, this necessitates the exploration of rigid translations parallel to the GB plane to adequately sample the different metastable states  \cite{sutton1983structure, VITEK1983,pond1980grain}. 
Over the past several decades, there have been numerous robust computational studies of GB structures and energies \cite{wolf1989correlation, Rittner110, tschopp2007, olmsted2009survey1, olmsted2009survey2, tschopp2015, ratanaphan2015grain, homer2022examination}. However, while most of these computational studies have considered $0$~K simulation conditions using Molecular Statics (MS), finite-temperature effects are well-known to strongly influence the mechanical properties of GBs \cite{rutter1965migration}.

To study GB properties under finite-temperature conditions, atomistic techniques such as Molecular Dynamics (MD) or Monte-Carlo (MC) sampling methods are commonly used. These have provided many insights into the effect of temperature on GB properties such as the complex temperature dependence of GB mobility  \cite{homer2014trends}. However, their computational cost and complexity compared to $0~\mathrm{K}$ MS limit the explorable space of GBs. Moreover, MD calculations of the free energy are complicated, as entropy is not directly accessible from the generated atomic trajectories. 
Occasional finite-temperature atomistic simulations that compute GB entropy (including anharmonic effects) date back to the work of \citet{entropyYip}, who obtained a reference entropy using the vibrational spectrum from the velocity autocorrelation function at low temperatures (where the harmonic approximation (HA) is valid). To compute entropies at the desired higher levels of temperature, where anharmonic effects are important, thermodynamic integration (TI) was used, with the aforementioned reference state serving as the starting point of the thermodynamic path. 
The studies that have been conducted since then (see, e.g., \cite{toda2012molecular, nishitani2021finite, eich2018embedded, matsuura2023anharmonicity}) have computed finite-temperature free energies using the Frenkel-Ladd path \cite{Frenkel19843188} and an Einstein solid as the reference state. This methodology has also been applied to GB phase transformations \citet{freitasS5} by using non-equilibrium states instead of equilibrium ones along the Frenkel-Ladd path to improve the efficiency of the calculations \cite{freitasNETI, freitasS5}.

An alternative family of techniques used for investigating the thermodynamic properties of GBs are free energy relaxation methods. These methods are less accurate but computationally more accessible than MD. In these methods, the integration of the dynamical equations of motion is bypassed by approximating the free energy using a simplified vibrational density of states (DOS). Some of the techniques of this type include the Local Harmonic approximation (LH) \cite{lesar1989finite, najafabadi1990finite}, the Variational Gaussian (VG) approximation \cite{lesar1989finite} and the temperature-dependent interatomic forces introduced by \citet{sutton1989temperature}. 
Such techniques rely on the assumption that atoms are uncoupled harmonic oscillators, whose stiffness depends on the mean positions of atoms through the interatomic potential $V$. The free energy of the solid is then calculated by minimization with respect to the atomic mean positions and vibrational frequencies. The combination with spatial coarse-graining has also made such methods popular in the quasicontinuum framework \citet{dupuy2005finiteQC, tadmor2013finite}.
These techniques are accurate at low temperature but become unreliable at higher temperature or for complex crystalline defects.
Improvements have been reported (up to about half the melting temperature of a given potential) by reintroducing coupling, as in the Quasi-Harmonic approximation (QHA).
Although there has been some success using the QHA to investigate GB energies \cite{scheiber2020temperature, langenohl2022dual, tuchinda2023vibrational,foiles1994evaluation, Sutton1994Appraisal} (by approximating the potential energy of the system up to second order about the finite-temperature equilibrium atomic positions), the free energy expression is too complex to be numerically minimized.
Therefore, the QHA does not offer a practical free energy minimization alternative to the aforementioned techniques. 

The limitation of the state of the art in finite temperature GB calculations is computational efficiency.
Surveys of finite-temperature GB properties over the 5D GB orientation space demand a highly efficient computational methodology, due to the number of boundaries and the sometimes extremely large periodic cells necessary to resolve the boundaries.
This work aims to address this need by combining the QHA with a statistical mechanics-based framework for solving quasistatic atomic relaxation, referred to as the Gaussian Phase Packets (GPP) approximation~\cite{gpp2021}, to calculate GB properties at finite temperature efficiently. This enables the accelerated calculation of atomic positions along with increased accuracy in the calculation of relaxed atomic structures. The remainder of this work is structured as follows. In \Cref{methods}, the methodologies used in this work for computing GB free energy and quasistatic shear-coupling factors, using the GPP framework and the QHA, are presented. \Cref{results and discussion} summarizes the set of investigated GBs and reports the obtained free energies of the lowest-energy metastable states from 100-500~K together with an analysis of their shear coupled behavior. Finally, we summarize the highlights of the current work, the advantages and disadvantages of the GPP framework for the study of GB properties, and potential future applications in \cref{conclusions}.

\section{Methodology} \label{methods}
\begin{figure}
\centering
\begin{tabular}{ccc}
\includegraphics[scale=0.25]{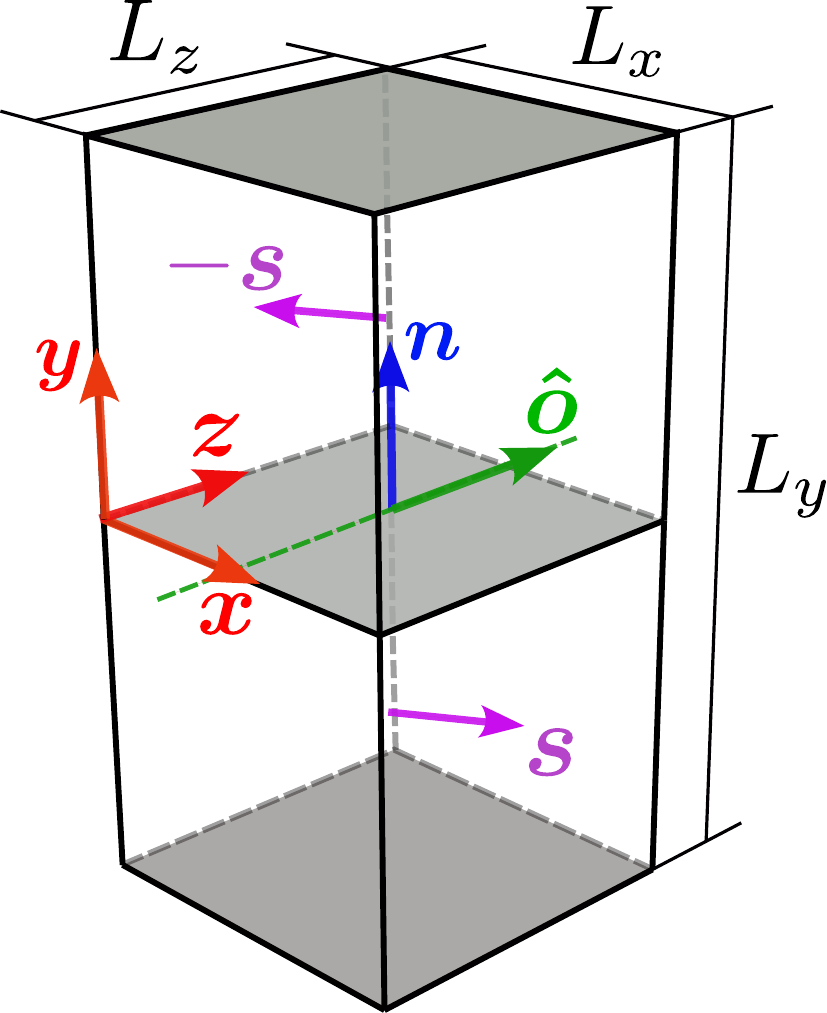} &
 \includegraphics[scale=0.25]{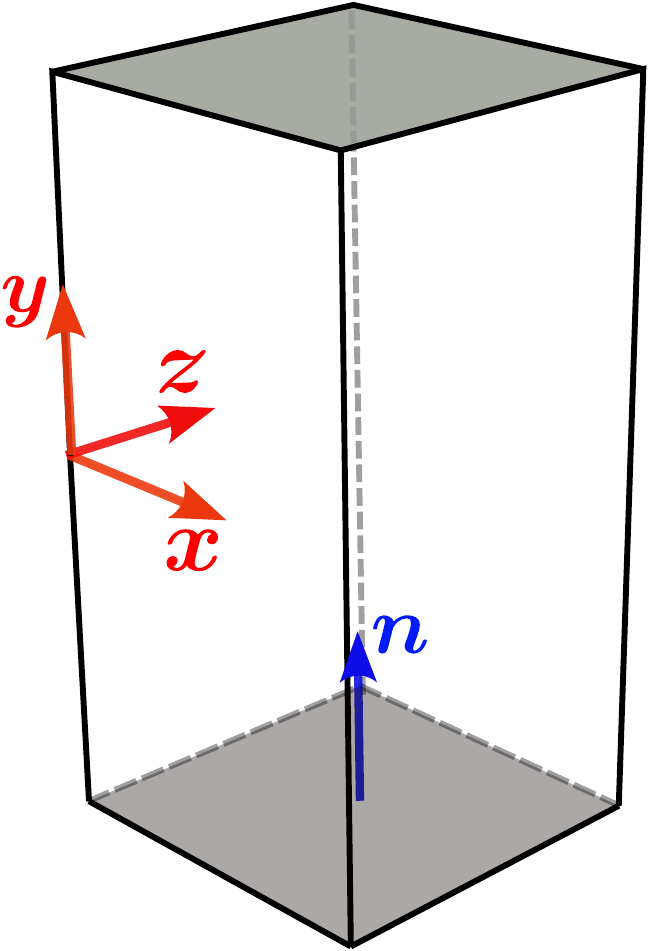} &
 \includegraphics[scale=0.25]{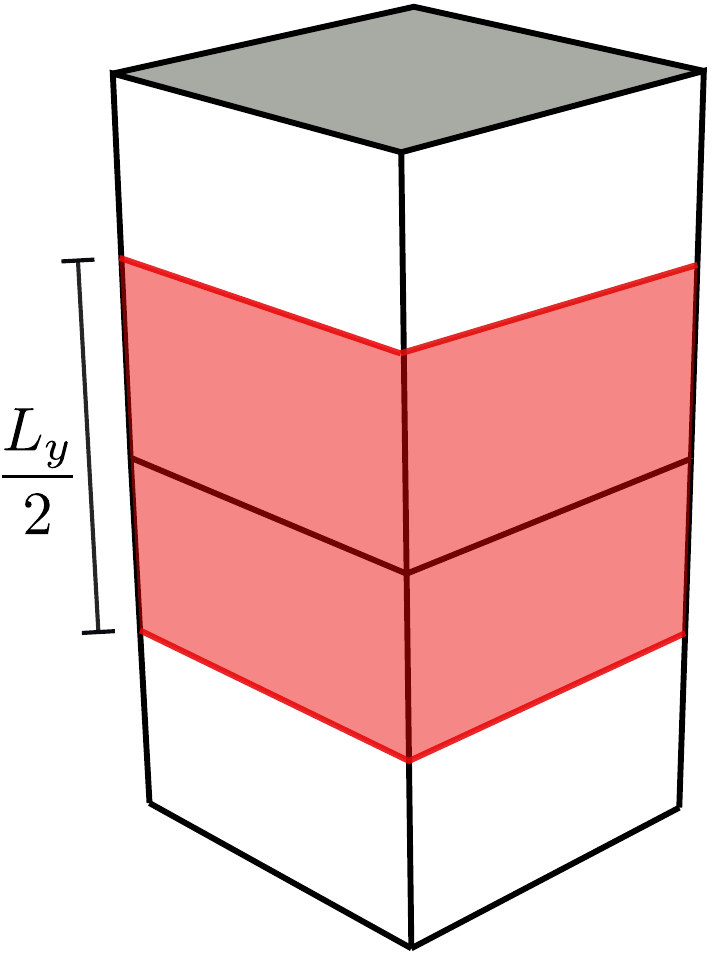} \\
(a) & (b) & (c)
\end{tabular}
\caption{Schematics of the different GB geometries used in this work, all with the same dimensions $L_x\times L_y\times L_z$. (a) Geometry of a bicrystalline sample with two free surfaces at the top and bottom, and the GB plane centered at $y=0$. (b) Geometry of a monocrystalline slab with the crystallographic orientation of one of the two grains and the dimensions of the bicrystalline sample. (c) Bicrystalline sample with red shaded regions representing the subset of atoms selected for the subsystem approach with $2d_\textsf{sub}=L_y/2$, where $L_y$ is the dimension of the samples in the $y$-direction.}
\label{fig:gpp_geom}
\end{figure}

This work presents finite-temperature calculations of the structure and free energy of a series of symmetric tilt grain boundaries (STGB) of face-centered cubic (FCC) Cu, modeled by the embedded-atom method (EAM) potential of~\citet{mishin2001potential} at temperatures of 100, 200, 300, 400, and 500~K. STGBs can be described by the relative orientation of the grains forming the GB and the crystallographic orientation of the GB plane normal $\boldsymbol{n}$. The relative orientation of the grains is specified by the angle $\theta$ by which one grain is rotated with respect to the other about the tilt axis $\hat{\boldsymbol{o}}$ (see \Cref{fig:gpp_geom}). We have considered a total of 19 values of $\theta$ for $\hat{\boldsymbol{o}}=\hkl[001]$, and 30 values of $\theta$ for $\hat{\boldsymbol{o}}=\hkl[011]$, each having a GB plane with a unique $\boldsymbol{n}$. 
We add two extra degrees of freedom $s_1$ and $s_2$ that define the Cartesian components (along the $x$- and $z$-axes, respectively) of the equal and opposite translation vectors $\boldsymbol{s}=[s_1, 0, s_2]$ and $-\boldsymbol{s}=[-s_1,0,-s_2]$, respectively, applied to each grain of the bicrystal parallel to the GB plane, see \Cref{fig:gpp_geom}a \citep{tschopp2007}. This allows the sampling of multiple metastable states for each $\{\boldsymbol{\hat{o}},\theta, \boldsymbol{n}\}$ triplet.  Additionally, for metastable states with minimum free energy in each set of metastable states with common triplet $\{\theta, \boldsymbol{\hat{o}}, \boldsymbol{n}\}$, we perform quasistatic displacement-driven shear (DDS) simulations, as outlined in \Cref{QSSC}. Further details on the sampled values of $\{s_1, s_2\}$ and specific geometric parameters used in the simulations will be presented along with the results in \Cref{results and discussion}.

\subsection{Grain boundary free-energy calculation}
\label{sec:GBEnergy}

The GB free energy $\gammaGB$ is defined as the excess Gibbs free energy per area induced by a GB in an otherwise defect-free crystal \cite{frolov2012thermodynamics2}. 
We consider stress-free grains, reducing the Gibbs to the Helmholtz free energy \cite{frolov2012thermodynamics1}. 
To compute this quantity, we consider a bicrystalline sample with the GB plane normal $\boldsymbol{n}$ aligned along the $y$-axis, GB tilt axis $\hat{\bfo}$ aligned along the $z$-axis, and two free surfaces parallel to the GB plane (\cref{fig:gpp_geom}a). 
To extract $\gammaGB$ we consider two methods that we will refer to as the `full-system' approach and the `subsystem' approach \cite{freitasS5}. 

The \textit{full-system} approach requires the absolute free energy of three geometries. The first geometry is the aforementioned bicrystalline sample. The free energy of this bicrystal will be denoted by $\mathcal{F}_\textsf{bi}$, the number of atoms it contains by $N_\textsf{bi}$, and the area of the GB by $A_\textsf{gb}$ (which is identical to the area of each free surface $(A_\textsf{surf})$). The free energy $\mathcal{F}_\textsf{bi}$ contains both the excess free energy introduced by the GB in the bulk material and the excess free energy introduced by the two free surfaces. We refer to this excess surface free energy per unit area as $\gammaSurf$. To compute the total surface free energy $2\gammaSurf A_\textsf{surf}$, we use a monocrystalline slab containing identical surfaces as the bicrystal, as shown in Figure~\ref{fig:gpp_geom}b. This system has free energy $\mathcal{F}_\textsf{slab}$ and $N_\text{slab}$ atoms. Lastly, computing $\gammaGB$ and $\gammaSurf$ requires the energy of atoms in the bulk material. This is obtained by using a third, infinite (using fully periodic boundary conditions) monocrystalline sample, having a free energy $\mathcal{F}_\textsf{bulk}$ and $N_\textsf{bulk}$ atoms, so the free energy per atom is simply $f_\textsf{bulk}=\mathcal{F}_\textsf{bulk}/N_\textsf{bulk}$. Having computed the free energy of these three systems, the bulk and surface free energy contributions are subtracted from $\mathcal{F}_\textsf{bi}$ to yield the GB free energy as
\begin{align}
\gammaGBfull=\frac{\mathcal{F}_\text{bi} - N_\text{bi}f_\text{bulk}}{A_\text{gb}} - 2\gammaSurf, \qquad \text{where} \qquad 
\gammaSurf=\frac{\mathcal{F}_\text{slab} - N_\text{slab}f_\text{bulk}}{2A_\text{gb}}.
\label{eqF_gb_gpp}
\end{align}

The \textit{subsystem} approach is an efficient approximation that only requires the first two geometries of the previous approach, namely the bicrystalline and monocrystalline slabs. In this approach we compute the free energy of the subset of atoms of the bicrystalline sample located a distance smaller than $d_\textsf{sub}$ from the GB plane (highlighted in red in Figure~\ref{fig:gpp_geom}c). 
We denote the free energy and the number of atoms of the subset by $\mathcal{\tilde{F}}_\textsf{bi}$ and $\tilde{N}_\textsf{bi}$, respectively. $\mathcal{\tilde{F}}_\textsf{bi}$ contains the full contribution of the GB and none from $\gammaSurf$ for appropriately chosen $d_\textsf{sub}$ and sample dimensions, which are detailed and justified in Section~3. Thus, by subtracting the free energy of the bulk from $\mathcal{\tilde{F}}_\textsf{bi}$, we can directly extract the excess energy introduced by the GB, while bypassing the computation of the surface free energy. To obtain the free energy of the bulk we select from the monocrystalline slab from the previous approach an analogous subset of atoms with the same dimensions as the subset selected in the bicrystal, as this subset is also free from $\gammaSurf$. We denote the free energy and the number of atoms of this subset $\mathcal{\tilde{F}}_\textsf{bulk}$ and $\tilde{N}_\textsf{bulk}$, respectively, and the bulk free energy per atom $\tilde{f}_\textsf{bulk}=\mathcal{\tilde{F}}_\textsf{bulk}/\tilde{N}_\textsf{bulk}$. The excess free energy per unit area of the GB obtained by this subsystem approach is referred to as $\gammaGBsub$ and given by
\begin{equation}
\gammaGBsub=\frac{\mathcal{\tilde{F}}_\textsf{bi} - \tilde{N}_\textsf{bi}\tilde{f}_\textsf{bulk}}{A_\textsf{gb}}.
\label{eqF_gb_gpp_sub}
\end{equation}
Due to the higher efficiency of the subsystem approach, we use it for all GB free energy calculations in this work. 
The full-system approach is only used in the convergence analysis in Section~\ref{conv analysis}, where both approaches are compared for validation.

On all samples we impose periodic boundary conditions in the $x$- and $z$-directions and free boundary conditions in the $y$-direction for the bicrystalline and monocrystalline slabs. The bulk sample has full periodic boundary conditions and the same crystallographic orientation as the monocrystalline slab and one of the grains of the bicrystal. The three samples were chosen to have the same dimensions $L_x$, $L_y$, and $L_z$, which were selected according to a convergence analysis (which will be presented in Section~\ref{conv analysis}). Identical dimensions between the samples ensure that the existing phonon modes in each sample are constrained to the same range of wavelengths. (A bulk sample larger than the bicrystalline sample includes bulk phonon modes absent in the bicrystal sample, so that the larger-wavelength bulk modes not captured in the bicrystal would incorrectly contribute to the GB excess entropy.) Keeping the same dimensions when computing the spring constants for an Einstein solid, as explained in \cite{freitasNETI}, follows the same argument. As a result of maintaining the same dimensions, the converge to $\gammaGB$ with sample dimensions is faster than when using an independently converged $f_\textsf{bulk}$, as shown in the convergence analysis in Section~\ref{conv analysis}.

Prior to applying any of the above two methods, we must ensure that the atomic structures of the samples correspond to a local minimum of the free energy. As an efficient and accurate technique for computing such equilibrium structures, referred to as \textit{relaxed} structures hereafter, we here use the GPP framework of \citet{gpp2021}, which is summarized in the following.

\subsection{Finite-temperature atomic structure relaxation via Gaussian Phase Packets}

In this section, we review the Gaussian Phase Packet (GPP) framework to the extent relevant to this work. For a detailed explanation of the approach, the reader is referred to \citet{gpp2021}. The GPP framework is used in this work to find atomic configurations with minimum free energy (i.e., the relaxed atomic structures). 

Consider an atomic ensemble of $N$ atoms defined by their positions $\bfq=\{\bfq_i(t) : i=1,\ldots,N\}$ and momenta  $\bfp=\{\bfp_i(t) : i=1,\ldots,N\}$ at time $t$. Using the condensed representation $\bfz=\left(\bfp(t),\bfq(t)\right)  \in \mathbb{R}^{6N}$ of the phase-space coordinate, the GPP framework defines a probability distribution function $f(\bfz,t)$, which quantifies the probability density of finding the atomic ensemble in a volume $\text{d}\bfz$ of the phase space at time $t$. More specifically it proposes a multivariate Gaussian for the probability distribution function, i.e.,
\begin{equation}
\label{eq:ansatz GPP}
    f(\bfz ,t) = \frac{1}{\calZ(t)} \exp\left[ -\frac{1}{2}\left(\bfz  - \bar{\bfz }(t)\right)\T \boldsymbol{\Sigma}^{-1}(t) \left(\bfz  - \bar{\bfz }(t)\right)\right],
\end{equation}
where $\bar{\bfz }(t)=
\langle\bfz\rangle=\int \bfz f(\bfz,t)\dd\bfz$ denotes the mean phase-space coordinate, and $\bfSigma\in \mathbb{R}^{6N\times 6N}$ is the covariance matrix of interatomic positions and momenta. $\calZ(t)$ is the partition function, defined via $\int f(\bfz,t)\dd \bfz = \langle 1\rangle = 1$, integrating over all of phase space (and denoting phase-space averages by $\langle\cdot\rangle$). Any thermodynamical quantity of interest can then be obtained as an integral of the observable weighted by the distribution function $f(\bfz,t)$ over the full phase space. Hence, determining the statistical parameters of the probability distribution function is the main objective of this approach as opposed to computing instantaneous degrees of freedom as in MD. The evolution equations for these statistical parameters can be obtained by inserting the above ansatz into Liouville's equation \cite{gibbs1902elementary}. Since solving for the evolution of all parameters in the covariance matrix $(\boldsymbol{\Sigma}(t))$ becomes intractable for large systems, we resort to an approximation of the probability distribution by assuming interatomic independence, which transforms the total probability distribution function into a product of individual Gaussian phase packets, according to
\begin{align}
     f(\bfz ,t) = \prod_{i=1}^N f_i(\bfz _i,t)
     \qquad\text{with}\qquad
     f_i(\bfz _i,t) = \frac{1}{\calZ _i(t)} \exp\left[ -\frac{1}{2}\left(\bfz _i - \bar{\bfz _i}(t)\right)\T \boldsymbol{\Sigma}_i^{-1}(t) \left(\bfz _i - \bar{\bfz _i}(t)\right)\right].
    \label{independent GPP}
\end{align}

For further simplification (and numerical tractability), a hyperspherical shape of the Gaussian distribution function is assumed for every atom. This simplifies the local covariance matrix $\boldsymbol{\Sigma}_i$ of atom $i$ into four diagonal matrices, such that
\begin{equation}
\label{GPP matrices}
\boldsymbol{\Sigma}^{(\bfp,\bfp)}_{i}=
\begin{bmatrix}
\Omega_i & 0 & 0 \\
0 & \Omega_i & 0 \\
0 & 0 & \Omega_i \\
\end{bmatrix}, \qquad
\boldsymbol{\Sigma}^{(\bfq,\bfq)}_{i}=
\begin{bmatrix}
\Sigma_i & 0 & 0 \\
0 & \Sigma_i & 0 \\
0 & 0 & \Sigma_i \\
\end{bmatrix},\qquad\text{and}\quad
\boldsymbol{\Sigma}^{(\bfp,\bfq)}_{i}= \boldsymbol{\Sigma}^{(\bfq,\bfp)}_{i}=
\begin{bmatrix}
\beta_i & 0 & 0 \\
0 & \beta_i & 0 \\
0 & 0 & \beta_i \\
\end{bmatrix}.
\end{equation}
The set of parameters to solve for every atomic site now reduces to $\{\bar{\bfp}_i, \bar{\bfq}_i,  \Omega_i, \Sigma_i, \beta_i  \}$. Since we are interested in the thermodynamic equilibrium configuration of the atomic ensemble, we use the quasistatic limit of the evolution equations for the statistical parameters. This leads to a vanishing mean momentum $\bar{\bfp}_i$ and mean thermal momentum $\beta_i$ for every atom. The equilibrium equations left to solve at every atomic site are the vanishing of the mean physical and thermal forces, which implies
\begin{equation}\label{EOM quasistatic}
\begin{split}
&\langle \itbf{F}_i \rangle = \itbf{0}\qquad \mathrm{and} \\
&\frac{\Omega_i}{m_i} + \frac{ \langle \itbf{F}_i(\bfq ) \cdot (\bfq_i  - \bar{\bfq}_i)  \rangle }{3} = 0,
\end{split}
\end{equation}
where $m_i$ denotes the mass of atom $i$, $\bfF_i$ is the net force on atom $i$.
The solution of \cref{EOM quasistatic} is the set of mean positions $\bar{\bfq}=\{\bar{\bfq}_{i} : i=1,\ldots,N\}$ and position variances $\Sigma=\{\Sigma_{i} : i=1,\ldots,N\}$ for all atoms. This solution may be re-interpreted as the minimizer of the free energy of the system \cite{gpp2021}. The momentum variance $\Omega_i$ is obtained by a-priori knowledge of the thermodynamic process involved in equilibriating the system. In this work, we are interested in equilibrium states of a canonical ensemble, and thus we impose isothermal conditions by setting $\Omega_i = m_i k_B T$ for every atom \cite{gpp2021}, where $T$ is the temperature of the ensemble. For numerical computations, phase averages $\langle \cdot \rangle$ in \cref{EOM quasistatic} are computed using Gaussian quadrature, while iterative relaxation is performed using the Fast Inertial Relaxation Engine (FIRE) \cite{fire}, starting from an initial guess $\{\Sigma_{i}, \boldsymbol{\bar{q}}_{i}\}$. Note that, in the quasistatic limit, this becomes equivalent to the max-ent \cite{kulkarni2008variational}, DMD \cite{li2011diffusive}, and variational Gaussian (VG) \cite{lesar1989finite} frameworks. 

The main difference between this approach and the local harmonic (LH) approximation \cite{lesar1989finite} is that the position variances here are allowed to relax independently from the mean positions, whereas the vibrational frequencies in the LH approximation are computed as the eigenvalues of the local dynamical matrices at every relaxation step. 
The GPP approach hence avoids the need for computing third-order derivatives and provides an additional degree of freedom per atom for the system to reach equilibrium. This approach has been shown to yield computationally efficient and accurate finite-temperature predictions of surface energies and elastic parameters \cite{saxena2022fast} as well as free energy variations across phase transitions \cite{saxena2023gnn} when compared to MD. However, it is important to note that the interatomic independence assumption (or local harmonicity) destroys all information about the complete phononic vibrational spectrum of the solid. \citet{foiles1994evaluation} showed that, while this assumption yields a fairly accurate prediction of relaxed atomic mean positions, absolute free energy values in the case of defects can be inaccurate \cite{rickman1993modified}. This is because the free energy is directly proportional to the system's entropy, which is sensitive to the eigenvalues of the dynamical matrix \cite{huang2022vibrational}. For a more detailed discussion, see Appendix~\ref{appendixA}. 

A more accurate alternative for free energy estimation, given a relaxed configuration pre-computed with GPP, is the fully coupled QHA (see Section~\ref{sec:QHA}).
This combined approach eliminates the need to assume interatomic independence (improving the free energy estimation) without the excessive cost of using QHA to find equilibrium structures.
In many works, the finite-temperature relaxed configuration used under the quasiharmonic approximation is considered to be simply a hydrostatic expansion of the 0~K relaxed configuration \cite{foiles1994evaluation, toda2012molecular, freitasS5}. However, this discards the possibility of individual atomic relaxations around defects as a function of temperature---which is of importance when considering GBs at finite temperature. Therefore, we pursue a two-step strategy in the following. We first use the above interatomic independent GPP formulation to solve for relaxed structures of GBs at finite temperature (in which anharmonic effects from the potential and local effects of crystalline disorder are taken into account) and then, in a post-processing step, use the QHA to recreate the complete dynamical matrix about the GPP-computed mean atomic positions and calculate the value of the free energy, as explained in the following. 

\subsection{Free energy calculation via quasi-harmonic approximation}
\label{sec:QHA}

The classical quasiharmonic approximation (QHA) relies on the assumption that the atomic displacements in a solid can be approximated as the superposition of the harmonic vibrational modes about their equilibrium positions $\bfq_0(T)=\{\bfq_{i0}(T) : i=1,\ldots,N\}$ at a temperature $T$. The potential energy of the ensemble is approximated by a Taylor series expansion up to second order, yielding
\begin{align}
    V(  \bfq   ) \approx V( \bfq_0  ) + ( \bfq_i - \bfq_{i0} )\T \frac{1}{2}\frac{\partial^2 V}{\partial \bfq_i \partial \bfq_j} \Bigg\vert_{\bfq_{i0},\bfq_{j0} } ( \bfq_j - \bfq_{j0} ).
    \label{eq:V_taylor}
\end{align}
The $3N$ vibrational frequencies $\omega =\{\omega_n : n=1,\ldots,3N\}$  and the normal vibrational modes, dependent on the equilibrium positions, can be obtained by diagonalizing the dynamical matrix (i.e., the mass-weighted force constant matrix)
\begin{align}
   \bfD = \sum \bfD_{ij} = \frac{1}{\sqrt{m_i m_j}} \frac{\partial^2 V}{\partial \bfq_i \partial \bfq_j} \Bigg\vert_{\bfq_{i0},\bfq_{j0} }.
\label{dynmat}
\end{align}
Once the vibrational frequencies are known, the free energy of the ensemble can be computed from the QHA as \cite{dove_1993}
\begin{align}
    \mathcal{F}_\text{QHA}(T) = V( \bfq_0 ) + k_BT\sum_{n=1}^{3N-3}\ln\left(\frac{\hbar \omega_n( \bfq_0 )}{k_BT}\right),
\label{eq:Fqha}
\end{align}
where we discard the three null eigenvalues corresponding to rigid-body modes ($k_B$ is Boltzmann's constant, and $\hbar=h/(2\pi)$ with Planck's constant $h$). It must be noted that under the QHA, the expansion in \cref{eq:V_taylor} is about the equilibrium positions $\boldsymbol{q}_0(T)$ at finite temperature $T$. In practice, obtaining $0~K$ equilibrium positions is trivial in comparison with their finite temperature counterparts. For that reason, the finite temperature positions for the QHA are usually approximated by hydrostatically expanding the $0~K$ equilibrium positions according to the bulk thermal expansion coefficient. As mentioned above, this discards the possibility of individual atomic relaxations at finite temperature, which results in inaccuracies. 
As a remedy, we use the GPP framework to efficiently obtain the minimum free energy structures and then use the QHA as a mere post-relaxation step to obtain accurate free energy values using the relaxed structures. Hence, the free energy, as required, e.g., in \cref{eqF_gb_gpp_sub}, is obtained as
\begin{equation}
\mathcal{F}(T) = V(\bar{\bfq}) + k_BT\sum_{n=1}^{3N-3}\ln\Bigg(\frac{\hbar \omega_n(\boldsymbol{\bar{q}})}{k_BT}\Bigg),
\label{qha_gpp}
\end{equation}
where $\bar{\bfq} $ is the set of mean atomic positions at finite temperature, which satisfies \cref{EOM quasistatic} (note the subtle but important difference between \cref{eq:Fqha,qha_gpp}). This allows us to exploit the computational efficiency of the GPP framework for obtaining accurate relaxed mean atomic positions at finite temperature with accurate free energy predictions using the complete vibrational spectrum about those relaxed positions as obtained from the QHA.

\subsection{Quasistatic shear coupling analysis} \label{QSSC}

Plasticity mediated by grain boundaries is increasingly understood to be a dominant deformation mechanism in many systems, including nanocrystalline (e.g. \cite{shan2004grain,jin2004direct}) and thin-film materials (e.g. \cite{mompiou2013inter}).
A key GB behavior that accommodates this phenomenon is shear-coupling, i.e. the shearing of a region swept by a moving boundary.
A key property determining a GB's shear-coupling behavior is the shear coupling factor, $\beta$, which can be determined quasistatically at 0K using traditional atomistics \cite{chesser2021optimal}.
Here, the quasistatic setting of GPP enables the determination of the shear coupling factor at finite temperature.

\begin{figure}[h!]
    \centering
    \begin{tabular}{c}
        \includegraphics[scale=0.2]{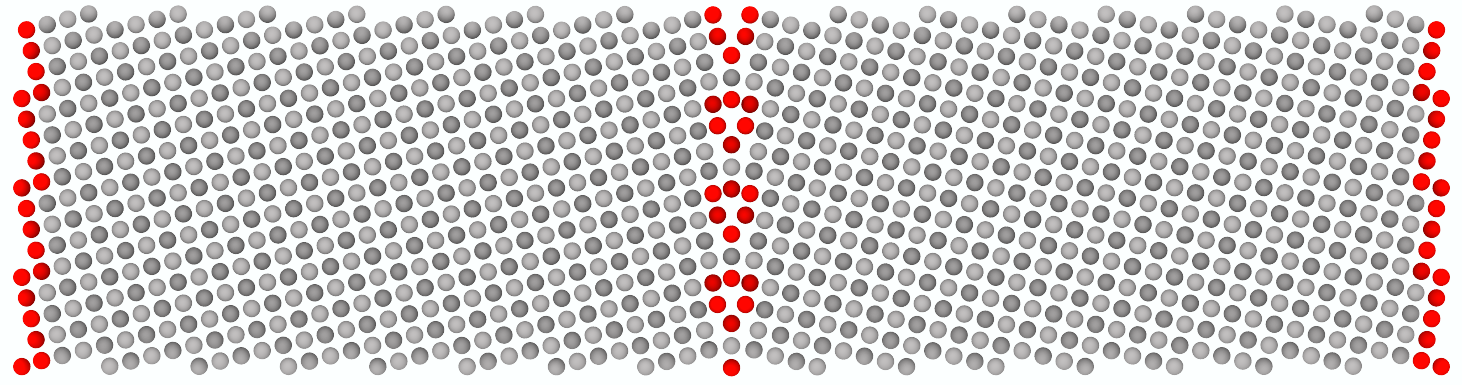} \\
        \includegraphics[scale=0.38]{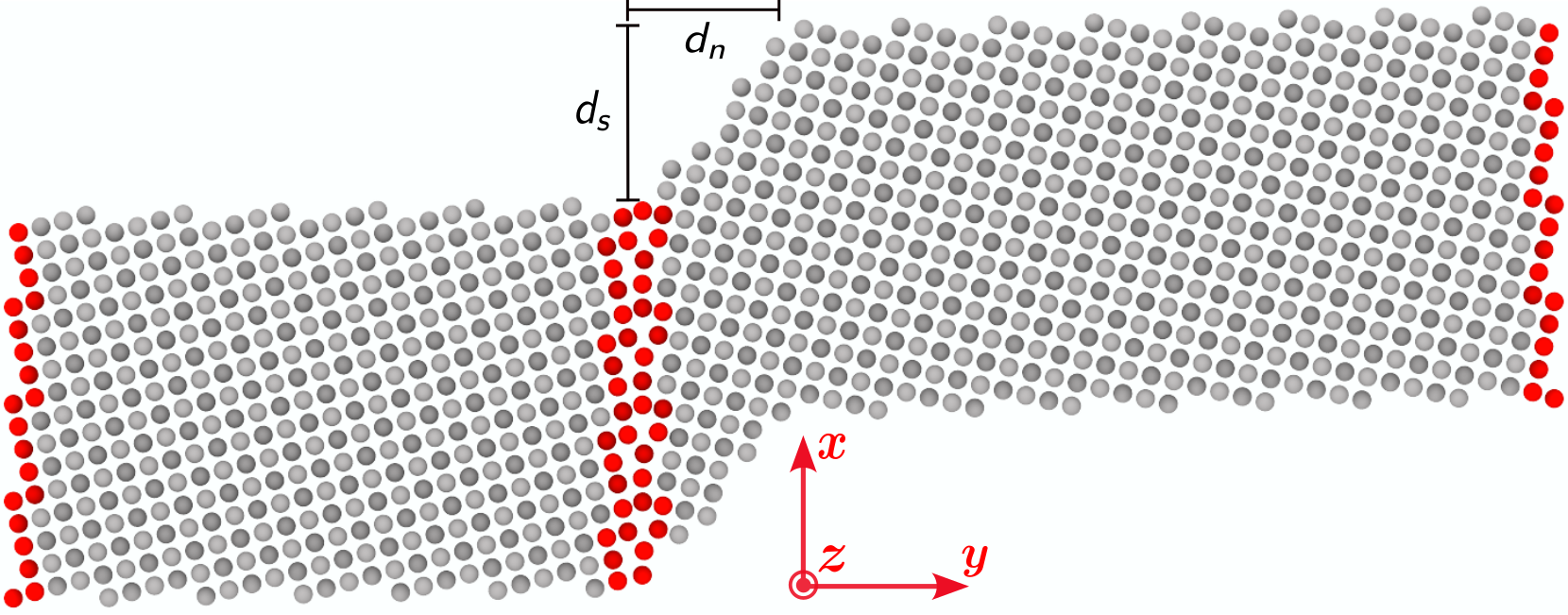}
    \end{tabular}
    \caption{Initial (top) and intermediate (bottom) atomic structures of a bicrystalline sample containing a $\Sigma 17$ GB in a quasistatic DDS simulation at $300~K$.} 
    \label{fig:gb_position}
\end{figure}

To this end, we subject the different equilibrated GBs configurations at each temperature $T$ to a quasistatic, displacement-driven shear (DDS) loading. To realize this in practice, we first select two slabs of atoms at the free surfaces of the bicrystal, hereafter referred as the `driving atoms'. The DDS simulation is then performed by displacing both slabs in opposite directions along the $x-$axis in increments of $0.1\si{\angstrom}$ each. After each displacement increment, a relaxation is performed to solve \cref{EOM quasistatic} for all atoms except for the driving atoms, whose degrees of freedom are kept fixed during this relaxation. This process is repeated for 100 iterations achieving a total shear displacement of $20\si{\angstrom}$.
The \textit{shear coupling factor} is given by 
\begin{equation}
    \beta={d_s}/{d_n},
\end{equation} 
where $d_s$ is the total shear displacement and $d_n$ the migration distance of the GB in the normal direction, as displayed in Figure~\ref{fig:gb_position}. To extract $\beta$ from the DDS simulations, we track the position of the GB plane by performing common neighbor analysis (CNA) in OVITO \cite{stukowski2009visualization}. We then perform a linear fit from the first slip of the GB plane and obtain $\beta$ as the slope of the fitted function.

\section{Results and discussion} \label{results and discussion}
Over $5{,}000$ initial bicrystalline geometries and the corresponding monocrystalline slabs, each corresponding to a unique set $\{\theta, \boldsymbol{\hat{o}}, \boldsymbol{n}, s_1, s_2\}$, were structurally relaxed using GPP at temperatures of 100, 200, 300, 400 and 500~K. To create these initial structures, we used a modified version of the script of \citet{gbscriptKIM}. The in-plane translational degrees of freedom $\{s_1,s_2\}$ along the $x$- and $z$-directions range from $0$ to $0.5\left[p_1(T)-0.015a(T)\right]$ and $0$ to $0.5\left[p_2(T)-0.015a(T)\right]$, respectively, in steps of $0.015a(T)$. Lattice spacing $a(T)$ is the temperature-dependent bulk atomic spacing, which was set to $a_\textsf{GPP}$ from Table~\ref{tab:lattice_spacings}, while $p_1(T)$ and $p_2(T)$ denote the periods of the CSL lattice in the GB plane directions $x$ and $z$, respectively. 
The bulk atomic spacing $a_\textsf{GPP}(T)$ was determined by solving \cref{EOM quasistatic} for a cube of $16\times16\times16$ FCC unit cells with periodic boundary conditions in every direction at temperature $T$ with the constraint of zero hydrostatic mechanical pressure. As shown by \citet{saxena2023gnn}, the inobjective nature of the numerical quadrature used to compute phase averages in \cref{EOM quasistatic} leads to slight variations in the relaxed lattice spacing values for different initial lattice orientations of the bulk. Since the quadrature perturbations in this work are always chosen to be aligned with the Cartesian directions, monocrystalline slabs with the same orientation as the bicrystal are used and relaxed to match the bulk lattice spacing in both samples and thus to improve the accuracy when computing the GB energy. Lastly, when setting up the initial GBs, one of the atoms in every pair of atoms closer than $d_\text{del}=0.5a(T)$ is deleted to avoid atoms being too close for relaxation. The critical value was selected, as it was verified to identify minimum-energy metastable states with the same energies reported by \citet{cahn2006} at $0~K$.

To determine the computationally optimal size of the bicrystalline samples (sufficiently large for accuracy but sufficiently small for numerical efficiency), we performed a convergence analysis to assess the influence of $L_x$, $L_y$, and $L_z$ on $\gammaGBsub$ and $\gammaGBfull$ of a $\Sigma5$ GB (details are summarized in Section~\ref{conv analysis}). Based on the convergence study, all samples were constructed with a GB plane area larger than $7.5~\mathrm{nm}^2$ with $L_z=32.5872~\si{\angstrom}$ for the cases with $\hat{\bfo}=\hkl[001]$ and $L_z=30.7235~\si{\angstrom}$ for $\hat{\bfo}=\hkl[011]$ at $100~K$. At higher temperatures, these dimensions were expanded according to the bulk thermal expansion predicted by the GPP framework (Table~\ref{tab:lattice_spacings}). The distance between the GB plane and the free surfaces in the bicrystalline samples was at least $40~\si{\angstrom}$, so $L_y\geq80~\si{\angstrom}$ in all initial geometries. Moreover, the distance used to choose the subset of atoms needed in the subsystem method (detailed in Section~\ref{methods}) was $d_\textsf{sub}=L_y/4$ (see Figure~\ref{fig:gpp_geom}). Lastly, as explained in Section~\ref{methods}, we compute free energies of stress-free grains, so that the periodic dimensions of the bicrystalline and monocrystalline geometries along the $x$- and $z$-axes were constant during relaxation. Following \citet{freitasS5}, this is equivalent to computing $\gammaGB$ in a stress-free, infinite bulk, thus avoiding grain size effects. Investigating the stress dependence of the GB free energy, as reported by \citet{frolov2012thermodynamics1, frolov2012thermodynamics2}, is a potential further step, which lies outside the scope of this work.

\begin{table}
\centering
\begin{tabular}{c|c|c|c|c|c}
     $T (K)  $ & 100 & 200 & 300 & 400 & 500  \\ \hline
     $a_\textsf{GPP}(\si{\angstrom})$ & 3.6208  & 3.6262  & 3.6315  & 3.6366  & 3.6416 \\ \hline
     $a_\textsf{MD}(\si{\angstrom})$ &  3.6205 & 3.6260  & 3.6316  & 3.6378   & 3.6442 \\
\end{tabular}
\caption{Bulk equilibrium lattice spacings at different temperatures, obtained with GPP as well as from a zero-pressure, isothermal ensemble using MD.}
\label{tab:lattice_spacings}
\end{table}

\subsection{Grain boundary free energies and structures}
\begin{figure}
    \centering
    \includegraphics[scale=0.65]{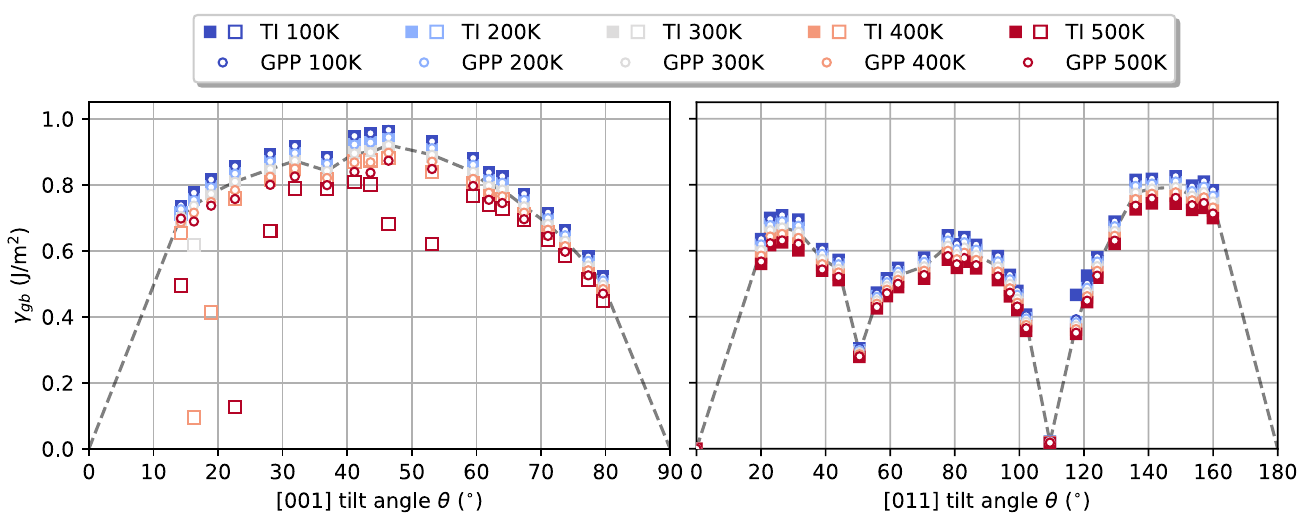}
    \caption{Free energies of selected metastable states with lowest $\gammaGB$ as obtained from GPP vs.\ from TI. Squares represent the average value of 5 TI runs with error bars extending from minimum to maximum values among the 5 runs. Open squares represent energy values obtained from TI runs (some atoms diffused further than $a^2/2$, so that the non-equilibrium Frenkel-Ladd path does not provide correct data).}
    \label{fig:gammas_gb}
\end{figure}

After relaxing the initial geometries and computing their free energies using \cref{qha_gpp}, we select for each GB configuration those metastable states with the lowest GB free energy $\gammaGB$. 
The GB free energies of these metastable states are shown in Figure~\ref{fig:gammas_gb} for the temperature range from 100 to 500~K. For comparison, the initial geometries yielding these selected metastable states were subjected to non-equilibrium thermodynamic integration (TI) \cite{freitasNETI} along the Frenkel-Ladd path, and the resulting GB free energies are also shown in Figure~\ref{fig:gammas_gb}. (For a description of the Frenkel-Ladd TI method and details of the simulation setup, see \cref{appendixB}.) Overall, the GPP-relaxed structures in combination with free energy determination by QHA show good agreement with the TI results---in all those cases were the latter yield reasonable results. TI simulations, especially at elevated temperature, are less reliable for the following reason (as reflected in the reported TI data in Figure~\ref{fig:gammas_gb}).

For the non-equilibrium Frenkel-Ladd TI method to provide reliable data, energy dissipation errors introduced by the non-equilibrium nature of the method must be cancelled by combining the forward and backward paths.
However, atomic diffusion can break this condition, introducing errors in the TI data \cite{freitasNETI}. During TI simulations, the atomic mean squared displacement (MSD) was collected for every atom to extract a measure of atomic diffusion. 
Cases with at least one atom having a MSD greater than the squared nearest-neighbor distance are shown in Figure~\ref{fig:gammas_gb} as open square markers, whereas all other TI data are shown by solid square markers. 
Most $\hkl[001]$ GBs at $400~K$ and above present at least one atom with a MSD larger than the squared nearest-neighbor distance (implying significant deviations from their initial positions), whereas none of the $\hkl[011]$ cases do. 

\begin{figure}
\centering
\begin{tabular}{cc}
     \includegraphics[scale=0.65]{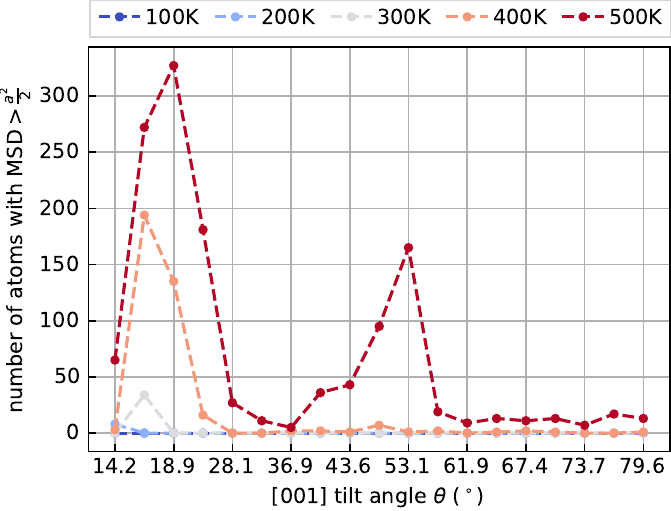} &
     \includegraphics[scale=0.65]{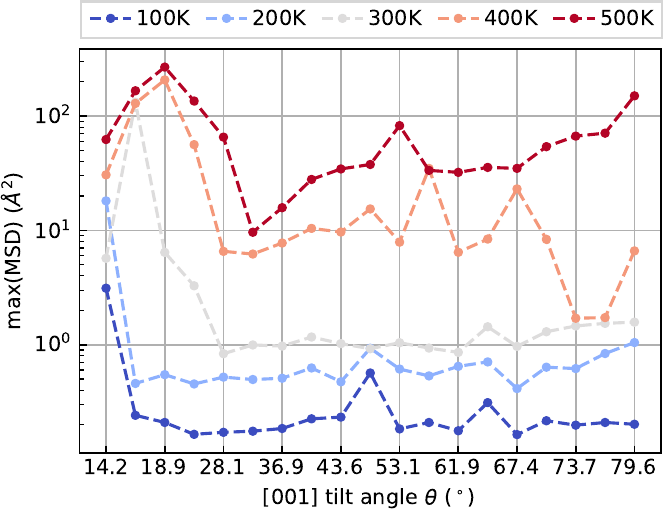} \\
     (a) & (b)
\end{tabular}
\caption{(a) Maximum number of atoms with mean squared displacement (MSD) larger than the FCC squared nearest-neighbor distance $a\sqrt{2}/{2}$, registered among 5 TI runs for each GB case with tilt axis $\hat{\bfo}=\hkl[001]$ at different temperatures. (b) Maximum MSD registered among 5 TI runs for each GB case with tilt axis $\hat{\bfo}=\hkl[001]$ at different temperatures. (The $\theta$-axes are not to scale for improved visibility.)}
\label{fig:msd}
\end{figure}

To further illustrate the degree of diffusion during the TI runs for each case, \cref{fig:msd} shows the number of atoms presenting MSDs larger than the squared nearest-neighbors distance for each case, together with the maximum atomic MSD recorded during the simulation.
At $400$~K, those cases with $\theta=16.26^\circ$, $\theta=18.93^\circ$ show one to two orders of magnitude more atoms with a MSD larger than the squared nearest-neighbor distance than the other cases. 
Moreover, at $400$~K they show maximum MSD values over $100\,\si{\angstrom}^2$. 
This indicates structural instability and hops of the system from one minimum to another in the potential energy landscape. Notably, out of the selected metastable states displayed in \cref{fig:gammas_gb,fig:msd}, only the two cases mentioned above showed at least one negative eigenvalue (excluding the three lowest ones, which correspond to rigid-body motion and are numerically $0$)\footnote{Note that negative eigenvalues here do not imply instability, as those are computed with the Hessian of the potential at $0$~K.}. More specifically, for $\theta=16.26^\circ$ negative eigenvalues were obtained at 100 K and higher temperatures, and for $\theta=18^\circ$ at 300 K and above. This overall clearly highlights the limitations of TI, especially at temperatures at and above $400$~K, where the presented GPP/QHA strategy still presents reliable data without numerical complications.


Additionally, as shown later in \cref{conv analysis}, our approach yields results more accurate than the classic QHA approach (see \cref{fig:convergence}). The average relative errors across the $\boldsymbol{\hat{o}}=\hkl[001]$ cases are approximately 0.02\%, 0.08\%, 0.15\%, and 0.38\% for 100, 200, 300 and 400~K, respectively, where the cases corresponding to $\theta=16^\circ$ and $18^\circ$ have been discarded due to structural instability and thus large errors in the TI data, as discussed above. Relative errors at 500~K are not reliable by the same argument, as most GBs show instabilities at this temperature. Similar errors across the $\boldsymbol{\hat{o}}=\hkl[011]$ cases are observed, approximately 0.2\%, 0.06\%, 0.14\%, 0.25\%, and 0.28\% for 100, 200, 300, 400 and 500~K respectively. The higher average relative error at 100~K compared to the $\hkl[001]$ counterpart is due to the higher relative errors of the cases with $\theta=117.56^\circ$ and $\theta=121^\circ$. When neglecting these two cases, the error decreases to 0.03\% at 100~K, which is comparable to the cases with tilt axis $\hkl[001]$.

\begin{figure}
  \centering
  \begin{tabular}{cc}
    \includegraphics[angle=90, scale=0.65]{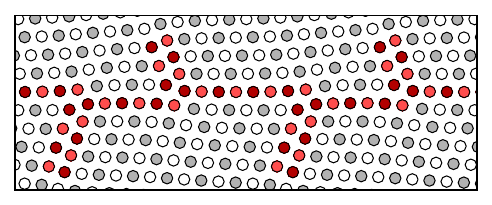} &
     \includegraphics[angle=90,scale=0.65]{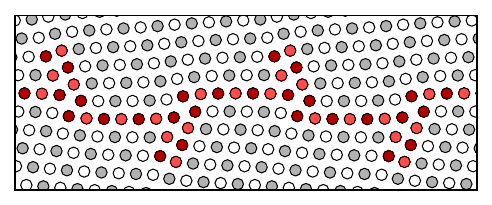} \\
    (a) 100~K & (b) 200~K
 \end{tabular}
 \caption{Two different $\Sigma67\hkl(-7-76)\hkl[011](\theta=117.56^\circ)$ metastable states, resulting from 0~K relaxation prior to TI from the same initial geometries (only differing by the lattice spacing $a_\textsf{MD}(T)$, as given in Table~\ref{tab:lattice_spacings}).}
\label{fig:gb_62_structure}
\end{figure}

\begin{figure}
\centering
\includegraphics[scale=0.65]{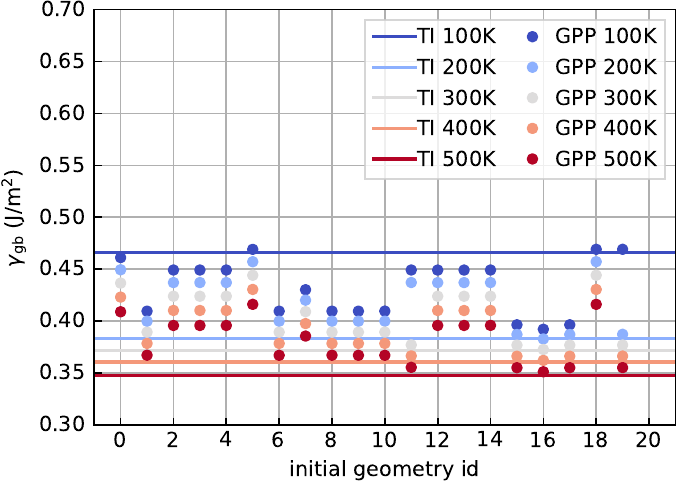}
\caption{$\gammaGB$ from twenty different initial geometries with $\{\boldsymbol{\hat{o}},\theta\}=\{\hkl[011],117.56^\circ\}$ relaxed with GPP and computed using \cref{qha_gpp} (circles) and $\gammaGB$ obtained from initial geometry with id=16 through TI (lines).}
\label{fig:gb_62_metastates}
\end{figure}

Inspection of the structures for the aforementioned cases of $\theta=117.56^\circ$ and $\theta=121^\circ$ exposes that the TI simulations in each case yield a higher-energy metastable state than that obtained from GPP at 100~K. As detailed in Appendix~\ref{appendixB}, TI is performed on the same initial geometries (with adjusted lattice spacing corresponding to that of MD) that lead to the metastable states with the lowest GB free energy using GPP (those shown in Figure~\ref{fig:gammas_gb}). Before the TI, a relaxation at 0~K is performed. For those two angles, this initial relaxation results in a different metastable state compared to that obtained from GPP for the same initial geometry at 100~K. As noted above, the initial geometries have different lattice spacings (and therefore in-plane displacements $s_1,s_2$) at different temperatures according to the bulk thermal expansion, potentially leading to different relaxed structures at different temperatures for equivalent initial geometry parameters. The two different metastable states for 100~K and 200~K from the TI simulations for the case with $\theta=117.56^\circ$ are shown in Figure~\ref{fig:gb_62_structure}. We also present the GB free energies obtained with GPP for a subset of the relaxed initial geometries with $\boldsymbol{\hat{o}}=\hkl[001]$ and $\theta=117.56^\circ$ in Figure~\ref{fig:gb_62_metastates}, which shows that some of the relaxed initial geometries at 100~K resulted in a metastable state with $\gammaGB$ close to that obtained from TI. More notably, some of these cases, specifically geometries with IDs 11 and 19, switched to a lower energy metastable state at higher temperatures as observed in the TI data. Hence, using the GPP for sampling metastable states can yield different GB metastable states when starting from the same initial geometry as temperature increases, thus demonstrating its ability to access a large number of GB structures and to capture shifts in minimum-energy GB configurations with temperature.

\subsection{Shear coupled motion}
The shear-coupled motion of those minimum-energy states shown in Figure~\ref{fig:gammas_gb} was studied at temperatures between 100 and 500~K. To this end, the GPP framework was used to perform quasistatic DDS simulations and compute their shear coupling factors $\beta$, following the methodology detailed in Section~\ref{methods}. 

\begin{figure}
    \centering
     \includegraphics[scale=0.65]{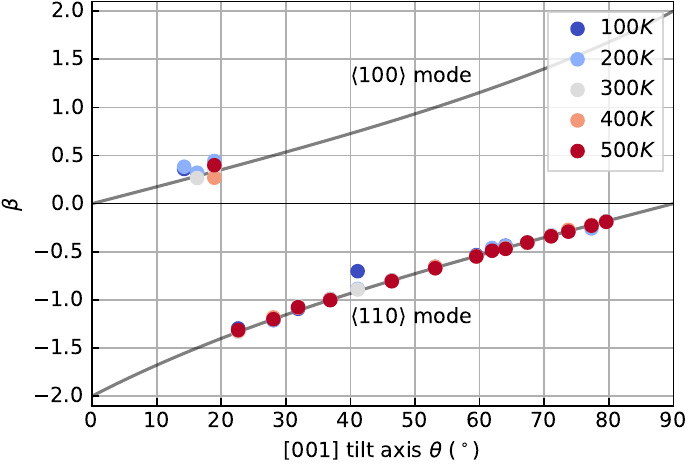}
    \caption{Computed shear coupling factor $\beta$ for different metastable states of $\hkl[001]$ tilt boundaries at different temperatures. Solid lines represent the analytical values of the two possible perfect coupling modes.}
    \label{fig:scf}
\end{figure}

Let us first discuss the data obtained for the $\hkl[001]$ tilt GBs, which are shown in Figure~\ref{fig:scf}. (Cases where $\beta$ cannot be resolved, such as in cases where the GB structure becomes unstable and diffuses into scattered defects impeding the identification of the GB position, are not included.) Most of the $\hkl[001]$ tilt GBs move under perfect coupling following one of the two modes predicted by the Frank-Bilby equation
\begin{equation}
\beta_{\langle 1 0 0 \rangle} = 2\tan\left(\frac{\theta}{2}\right) \qquad \text{and} \qquad \beta_{\langle 1 1 0 \rangle} = -2\tan\left(\frac{\pi}{4} - \frac{\theta}{2}\right)
\end{equation}
as reported by \citet{cahn2006}. While most GBs show perfect coupling in the $\langle110\rangle$-mode, lower-angle cases with $\theta=14.25^\circ,16.26^\circ,18.94^\circ$ show coupling values close to the perfect $\langle 100 \rangle$-mode. Deviations from perfect coupling in these three cases can be attributed to the loss of their initial perfect kite structure in favor of disordered structures as the shear displacement increased. This is in line with the observed structural instability that these metastable states presented during TI, as discussed above. Overall, these results agree partially with \cite{cahn2006}, as the value of the shear coupling factors for each mode are constant across temperatures and are thus a geometric factor. However, \citet{cahn2006} also observed cases with angles $\theta=22.62^\circ$ and $\theta=28.07^\circ$ to move in the $\langle 100 \rangle$-mode in MD simulations at 200~K and 400~K, respectively. The reason for the disagreement between the present work and the latter might be explained by the same argument proposed by \citet{cahn2006}, namely that the $\hkl<100>$ is accessible only by thermal activation of out-of-plane atomic movement. 

\Cref{fig:gb_position} (top) shows the kite structures highlighted in red in the center of the bicrystal, and \Cref{fig:gb_position} (bottom) shows the kite structures and the adjacent B structures also highlighted in red. For the $\langle 110 \rangle$-mode, B structures transform into adjacent kite structures and vice versa by in-plane (perpendicular to the tilt axis) atomic movements. These in-plane movements are driven by the applied shear displacement, and therefore the direction of the shear with respect to the orientation of the kite structures determines the migration direction. For the $\langle100\rangle$-mode, C structures \cite{cahn2006} transform into kite structures. For this mechanism to be possible, the out-of-plane symmetry of the lattice must be broken, which in \cite{cahn2006} was promoted by thermal fluctuations of the MD simulations. Nonetheless, in this work, we observed migration close to this mode in a quasistatic setting for the three lowest-angle cases, which became disordered with increasing shear displacement due to their structural instability, as discussed above. This instability and loss of perfect kite structure could be assisting the out-of-plane movements required for the $\langle 100 \rangle$-mode to unfold.

\begin{figure}
    \centering
    \begin{tabular}{ccc}
    \includegraphics[scale=0.45]{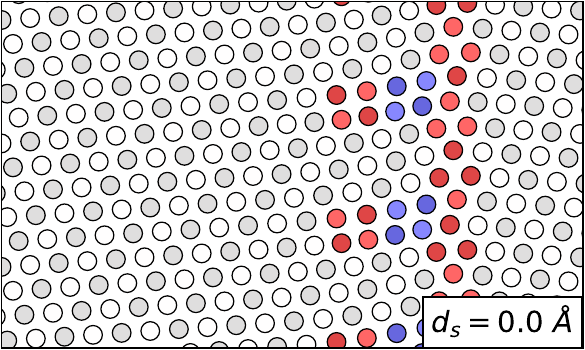} & \includegraphics[scale=0.45]{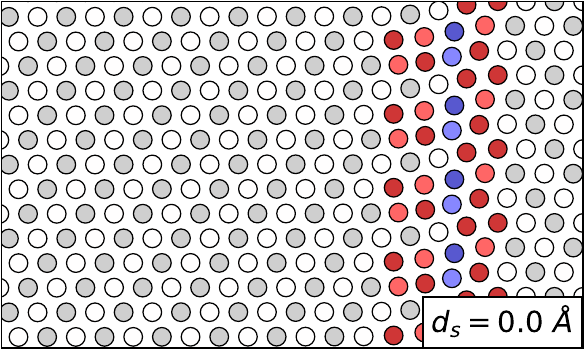} & \includegraphics[scale=0.45]{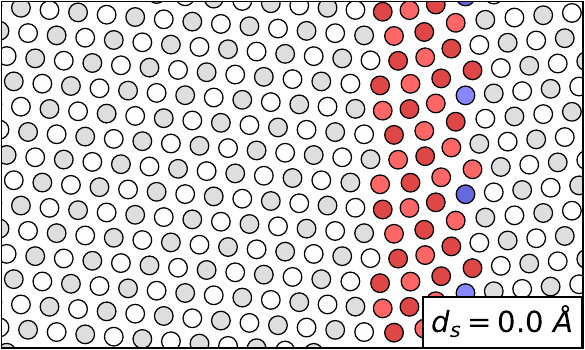} \\
    \includegraphics[scale=0.45]{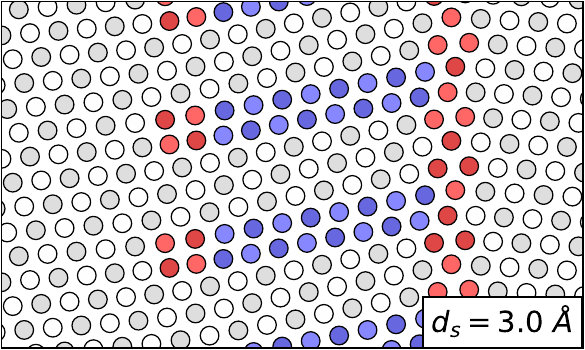} & \includegraphics[scale=0.45]{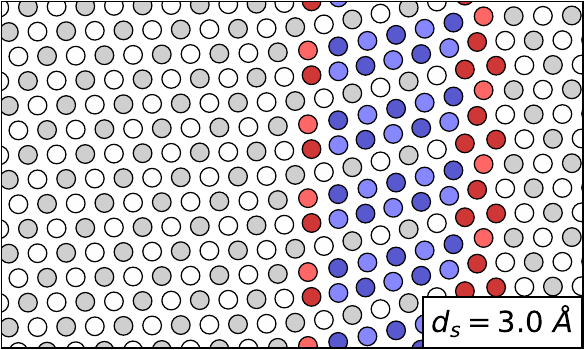} & \includegraphics[scale=0.45]{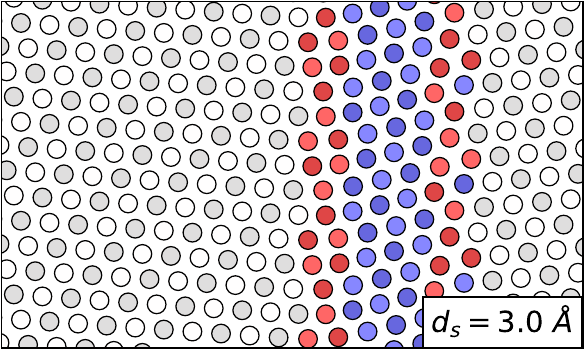} \\
    \includegraphics[scale=0.45]{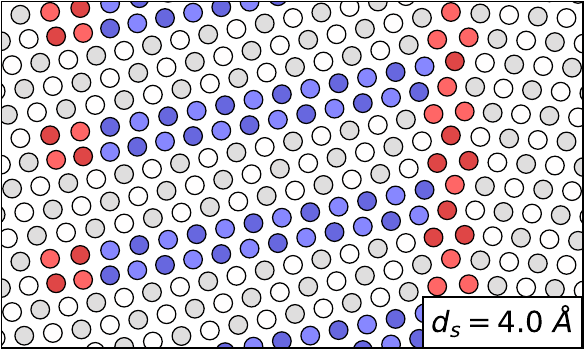} & \includegraphics[scale=0.45]{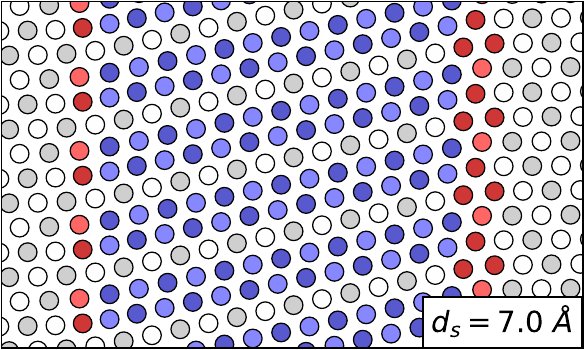} & \includegraphics[scale=0.45]{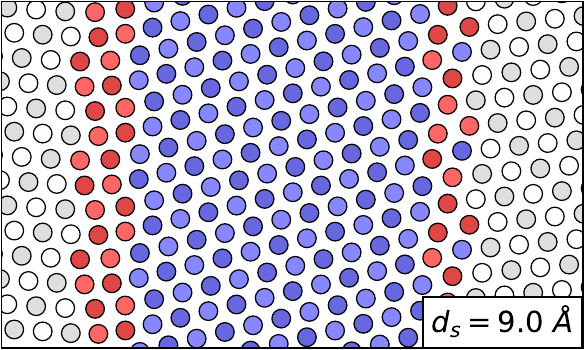} \\
        (a) & (b) & (c)
    \end{tabular}
\caption{GB structures during quasistatic DDS simulations at $300$~K for bicrystals containing (a) a $\Sigma67(\theta=62.4^\circ)$, (b) a $\Sigma3(\theta=70.5^\circ)$, and (c) a $\Sigma43(\theta=80.6^\circ)$ GB. Colors indicate crystal lattice structure according to common neighbor analysis: white for FCC, blue for HCP, red for other. Shading according to crystallographic planes with out-of-page normal.}
\label{fig:011_structs_a}
\end{figure}

\begin{figure}
\centering
\begin{tabular}{ccc}
\includegraphics[scale=0.45]{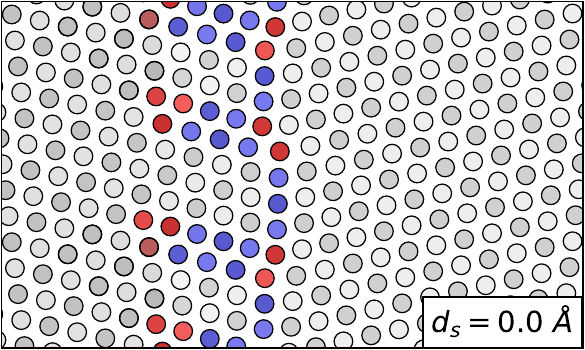} & \includegraphics[scale=0.45]{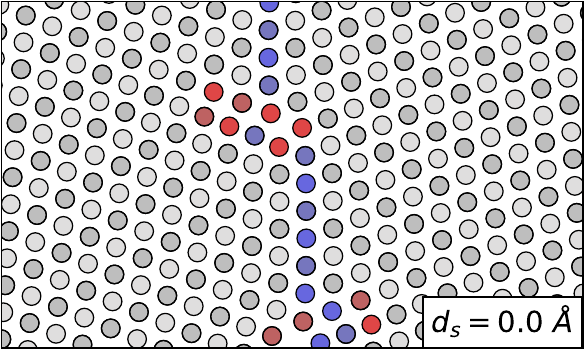} & \includegraphics[scale=0.45]{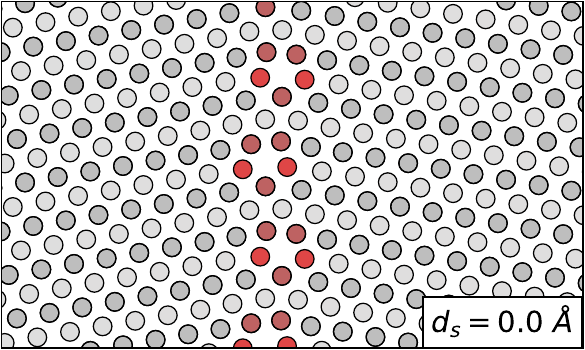} \\
\includegraphics[scale=0.45]{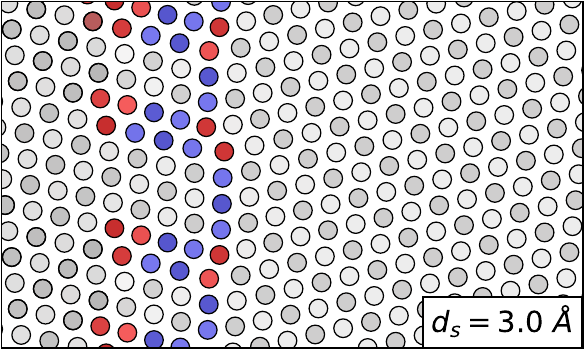} & \includegraphics[scale=0.45]{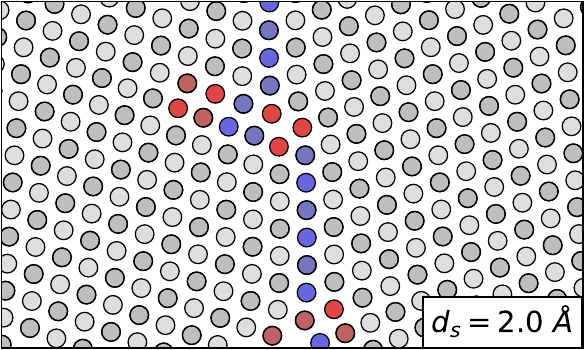} & \includegraphics[scale=0.45]{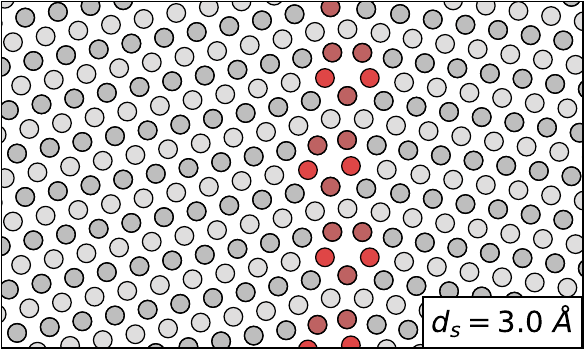} \\
\includegraphics[scale=0.45]{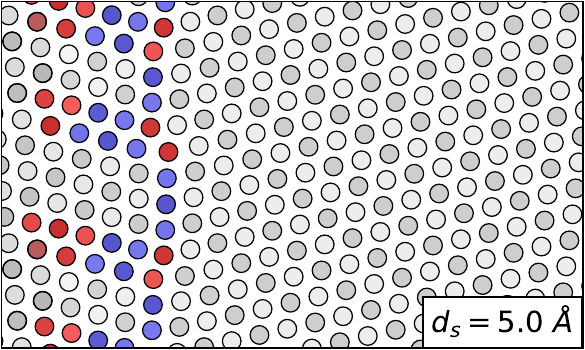} & \includegraphics[scale=0.45]{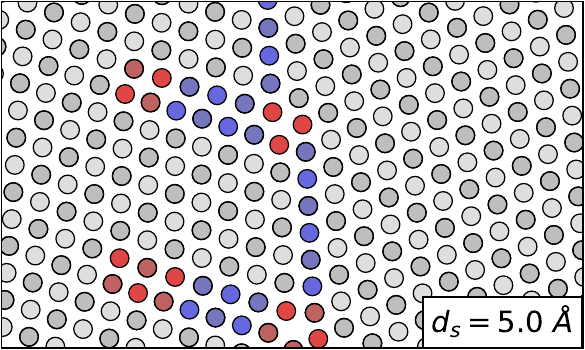} & \includegraphics[scale=0.45]{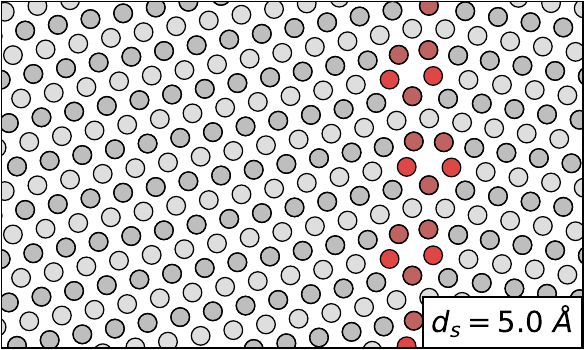} \\
        (a) & (b) & (c)
    \end{tabular}
\caption{GB structures during quasistatic DDS simulations at $300$~K for bicrystals containing (a) a $\Sigma17\hkl[011](\theta=93.34^\circ) $, (b) a $\Sigma67\hkl[011](\theta=117.56^\circ)$, and (c) a $\Sigma33\hkl[011](\theta=159.95^\circ)$ GB. Colors indicate crystal lattice structure according to common neighbor analysis: white for FCC, blue for HCP, red for other. Shading according to crystallographic planes with out-of-page normal.}
\label{fig:011_structs_b}
\end{figure}

In contrast to the $\hkl[001]$ tilt boundaries, most of the tested $\hkl[011]$ tilt boundaries do not present perfect coupling. \citet{homer2013phenomenology} also observed this general lack of coupling behavior in $\hkl[011]$ boundaries. Instead, a myriad of behaviors were observed. Notably, some generalities can be extracted for boundaries with similar $\hkl[011]$ tilt angle and free energy, as they share the same type of structural features. The most interesting behavior was found for intermediate $\hkl[011]$ tilt angles from $\theta=58.88^\circ$ to $\theta=102.12^\circ$. All these boundaries present stacking faults on one of the grains. As the tilt angle approaches $90^\circ$ from below, the distance between stacking faults becomes smaller, as shown in Figure~\ref{fig:011_structs_a}. Instead of preserving their structure, their stacking faults extended forming a new migrating boundary and nucleating a new phase between the grains. Specifically, the $\Sigma3\hkl(-1-12)\hkl[011](\theta=70.5^\circ)$ boundary grows a 9R phase, as has been observed experimentally for example in \cite{medlin1998stacking}, see Figure~\ref{fig:011_structs_a}b. Moreover, for the $\Sigma43\hkl(-3-32)\hkl[011](\theta=80.6^\circ)$ boundary the stacking faults become sufficiently close so that a hexagonal close packed (HCP) phase is formed in between, see Figure~\ref{fig:011_structs_a}c.  As the $\hkl[011]$ tilt angle approaches $90^\circ$ from above ($\theta=102.12^\circ$ to $\theta=93.37^\circ$), the distance between the stacking faults also decreases. These boundaries, in contrast, maintain their structure and undergo GB motion as a whole, see Figure~\ref{fig:011_structs_b}a. Higher $\hkl[011]$ tilt angle boundaries from $\theta=117.56^\circ$ to $\theta=124.12^\circ$ contained stacking faults in both grains in an alternating fashion, whose reach into the grain decreases with increasing angle. These boundaries were mostly immobile as the shear displacement progressively shifted all stacking faults to one of the grains, as depicted in Figure~\ref{fig:011_structs_b}b. On the high-angle end, a few GBs with high tilt angles presented stick-slip behavior. Specifically, boundaries $\Sigma9$ ($\theta=141.06^\circ$) and $\Sigma33$ ($\theta=159.95^\circ$), the latter presenting perfect coupling with $\beta\approx0.354$, see Figure~\ref{fig:011_structs_b}c). Boundary $\Sigma19$ ($\theta=153.5^\circ$) also presented stick-slip behavior, but it changed migration direction staying close to its initial position.

\subsection{Convergence analysis} \label{conv analysis}
To minimize size effect in the computed GB free energies and the presented shear coupling behavior reported above, the dimensions of the simulated samples were determined in a convergence analysis of the GB free energy with varying sample dimensions. Specifically, the convergence of $\gammaGBsub$ and $\gammaGBfull$ with dimensions $L_x,L_y,L_z$ for the full-system and subsystem approaches was performed for a $\Sigma5\hkl(310)\hkl[001]$ GB at $300$~K as a representative example with the objective of demonstrating convergence of both approaches to the same GB free energy and selecting the dimensions of the samples that yield sufficiently converged values. In this analysis, $L_y$ was assigned 3 values: $6p_{\hkl[310]}, 8p_{\hkl[310]}$ and $10p_{\hkl[310]}$, where $p_{\hkl[310]}$ is the period of the CSL lattice for the $\Sigma5\hkl(310)\hkl[001]$ boundary. Each of these 3 values was combined with four increasing values of the GB area $A_\text{gb}=L_x\times L_z$, totaling 12 sets of samples. Each set consists of a bicrystalline slab, a monocrystalline slab, and a bulk sample--- all with the same sizes and crystallographic orientations, as described in Section~\ref{methods}. For the full-system approach, all three samples are needed, while the subsystem approach requires only the first two. $\gammaGBfull$ and $\gammaGBsub$ were computed according to \cref{eqF_gb_gpp,eqF_gb_gpp_sub}, respectively. In addition, a modified version of the full-system method was explored by replacing $f_\text{bulk}$ in \cref{eqF_gb_gpp} by the per-atom free energy obtained from a larger sample of bulk material of $16\times16\times16$ FCC unit cells, denoted by $f_\text{bulk,lim}$. We refer to this approach as the `\textit{bulk limit approach}" and denote the excess free energy density obtained from it by $\gammaGBlim$.

\begin{figure}
    \centering
      \includegraphics[scale=0.65]{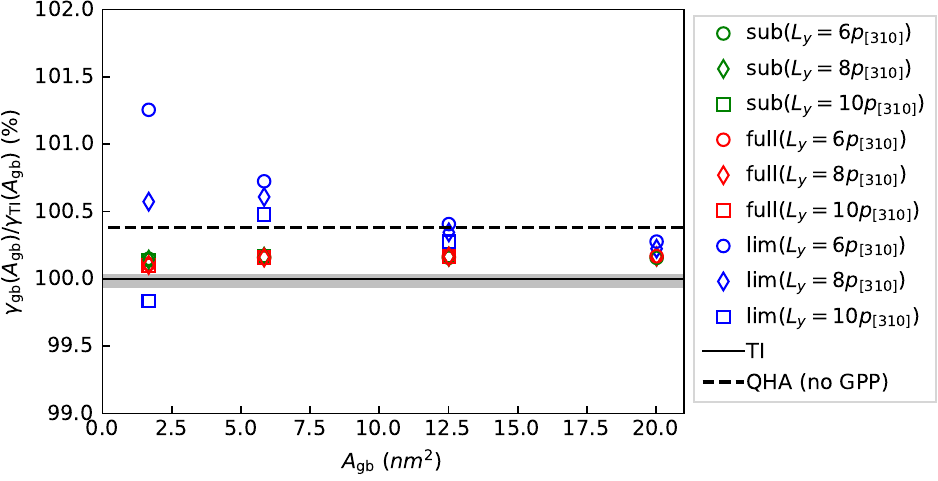}
    \caption{Convergence of $\gammaGB$ with $A_\text{gb}$ and $L_y$ for $\Sigma5\hkl(310)\hkl[001]$ at $300$~K. Red and green markers represent values obtained with the full-system approach (Eq.~\ref{eqF_gb_gpp}) and the subsystem approach (Eq.~\ref{eqF_gb_gpp_sub}), respectively. Blue markers represent results obtained from the bulk limit approach. The solid line denotes the average value from 5 TI runs, and the gray area extends from the minimum to the maximum values among these 5 runs. The dashed line represents the value obtained using the classical QHA, where the equilibrium positions were obtained by expanding the 0~K relaxed structure by the thermal expansion of the potential (Without the use of GPP).}
    \label{fig:convergence}
\end{figure} 

Figure~\ref{fig:convergence} shows the GB free energies obtained with each approach for each of the 12 sets. The subsystem approach shows the fastest convergence with $A_\text{gb}$ and $L_y$ among the three approaches. The full-system approach has similar convergence but requires the use of an extra sample and the computation of the dynamical matrix of a larger set of atoms. The bulk limit approach shows the lowest convergence rate, which may be explained as follows. In the latter approach, the subtracted bulk free energy includes more vibrational modes with larger wavelengths than the bicrystalline and monocrystalline samples due to its larger dimensions, while in the subsystem and full-system approaches the same dimensions of the bicrystal, monocrystal, and bulk samples result in vibrational modes limited to the same wavelengths for each sample. As a consequence, the bulk limit approach converges more slowly, as it requires larger bicrystal and monocrystalline samples with dimensions equivalent to the larger bulk sample used to compute the bulk free energy, so that the contributions of all bulk modes are cancelled. Based on this data, we selected the subsystem approach for all calculations, as its convergence is the fastest and its computational cost is lower than that of the full-system approach. Based on Figure~\ref{fig:convergence} and aiming for a good trade-off between sufficiently converged values and low computational cost, the dimensions of the samples were chosen as $A_\text{gb}\geq7.5nm^2$ and $L_y\geq8nm$ (which is larger than $6p_{\hkl[310]}$).

It is important to note that the convergence of $\gammaGB$ with sample size is different for GPP and MD simulations. The dimensions required for convergence in MD \cite{freitasS5} are typically larger than those obtained in this convergence analysis for GPP. Both techniques need samples sufficiently large to resolve the full elastic fields arising from the GB into the grains and to capture the significant phonons introduced by the GB. However, MD generally requires large system sizes to explore a sufficiently large fraction of phase space, which is critical for the entropic contributions to the free energy. (Moreover, statistical fluctuations in MD require several runs per TI simulation for averaging.) By contrast, the GPP formulation \cite{gpp2021} approximates phase averages by Gaussian quadrature, resulting in every equivalent cluster of atoms yielding the same thermodynamical properties without fluctuations, thus allowing for the use of significantly smaller samples (e.g., for bulk samples only a single atomic neighborhood) and hence accelerating the free energy computation.

\section{Conclusions} \label{conclusions}
In this work, we have studied the relaxed GB structures, free energies, and shear coupling factors of a series of $\hkl[001]$ and $\hkl[011]$ STGBs at finite temperature, using the quasistatic setting of the Gaussian Phase Packets (GPP) framework. GPP is a free-energy minimization framework that allows to perform computationally efficient structural relaxation at finite temperature. This enables a more accurate prediction of the relaxed GB structure at finite temperature compared to a simple hydrostatic expansion about the 0~K relaxed structure, as seen in previous works on GB free energies. Over $5{,}000$ bicrystals spanning a total of 49 different GBs have been relaxed using this approach, leading to the identification of different metastable states for each of these GBs as a function of temperature. Interestingly, some GB configurations showed shifts in metastable state (starting from the same initial bicrystalline geometries) with increasing temperature. The free energy of these metastable states has been obtained using the QHA as a post-relaxation step and, for comparison, using thermodynamic integration, yielding excellent agreement between the two. We have also computed shear coupling factors under quasistatic conditions through the GPP framework. The shear coupling factors of GBs with $\hkl[001]$ tilt axis have been correctly predicted for most boundaries, confirming that they do not depend on temperature and agreeing with results from \citet{cahn2006} and others. We have also confirmed the general lack of perfect shear coupling in GBs with $\hkl[011]$ tilt axis \cite{homer2013phenomenology}, reported general trends in their kinetics according to energy and structural features, and identified interesting cases where intermediate phases are nucleated, such as the 9R phase in $\Sigma3$ twin boundaries.

Overall, the GPP framework has proven to be a valuable tool for the study of GB properties at finite temperature, as it can sample larger sets of GB structures more efficiently than MD-based methods (see also \cite{saxena2022fast} for a discussion of the computational expenses of the GPP framework). The assumption of interatomic independence makes the method computationally efficient and accurate for structural relaxations but results in considerable errors when computing the vibrational entropies and therefore free energies for complex crystallographic defects such as GBs. This is the reason why the QHA has been used to restore the vibrational spectra as a post-processing step for the GPP finite-temperature structural relaxations. This provides accurate GB free energies at low to intermediate temperatures, but still suffers from inaccuracies at higher temperatures due to the harmonic nature of the method. However, it does significantly improve the results of QHA on hydrostatically expanded 0~K structures. The GPP framework has also proven valuable to investigating the shear coupled motion of GBs in a quasistatic setting, though one downside of this setting is that thermally activated processes are difficult to capture. Thus, two of the low-angle $\hkl[001]$ tilt GBs were not found to move under the thermally-activated $\langle 100\rangle$-mode as predicted by \citet{cahn2006}. Possible extensions with the introduction of stochasticity would extend its capability for capturing the impact of thermal fluctuations.  

\section*{Acknowledgement}
The support from the European Research Council (ERC) under the European Union’s Horizon 2020 research and innovation program (grant agreement no.~770754) is gratefully acknowledged.
BR acknowledges support from the United States National Science Foundation, grant number MOMS-2341922.
This work used the INCLINE cluster at the University of
Colorado Colorado Springs. INCLINE is supported by the National Science Foundation, grant number OAC-2017917. MS also thanks Rodrigo Freitas for his support in setting up the TI simulations and QHA free-energy calculations. 


\bibliographystyle{cas-model2-names}
\bibliography{grain_boundaries}

\appendix

\section{Vibrational entropies: GPP vs.\ QHA} \label{appendixA}
To illustrate the effect of discarding interatomic correlations in the GPP framework, we here compare the GPP and QHA vibrational spectra and vibrational entropies for two defects: an FCC $\Sigma5\hkl(310)\hkl[001]$ GB and a $\hkl[310]$ surface. First, consider the canonical probability distributions under the QHA and the GPP ansatz in \cref{eq:ansatz GPP}, both under isothermal equilibrium conditions (based on an NVT ensemble):
\begin{align}
f_\textsf{QHA} &=  
\frac{1}{\calZ_\textsf{QHA}}\exp\bigg[-\frac{1}{k_BT}\sum_{i=0}^{N}\frac{\bfp_i^2}{2m_i}\bigg]\exp\bigg[-\frac{1}{k_BT}( \bfq_i - \bfq_{i0} )^T \frac{1}{2}\frac{\partial^2 V}{\partial \bfq_i \partial \bfq_j} \Bigg\vert_{\bfq_{i0},\bfq_{j0} } ( \bfq_j - \bfq_{j0} )\bigg], \\
f_\textsf{GPP} &= \frac{1}{\calZ_\textsf{GPP}} \exp\left[ -\frac{1}{2}\sum_{i=0}^N\frac{\bfp_i^2}{\Omega_i}\right]\exp\left[-\frac{1}{2}(\bfq_i  - \bar{\bfq}_i)^T \boldsymbol{\left[\Sigma^{(\bfq,\bfq)}\right]}^{-1} (\bfq_j  - \bar{\bfq}_j)\right],
\end{align}
where $f_\textsf{GPP}$ is a simplified version of Eq.~\ref{eq:ansatz GPP} with vanishing mean momenta $\bar{\bfp}\rightarrow 0$ and momentum-position correlations $\beta\rightarrow 0$ due to equilibrium. Consequently, under the aforementioned conditions, the dynamical matrix $\bfD$ containing the Hessian of the potential energy (see \cref{dynmat}) and ${\Sigma^{(\bfq,\bfq)}}^{-1}$ containing the position-position variances, are related by
\begin{equation}
\frac{\left[\boldsymbol{\Sigma}^{(\bfq,\bfq)}\right]^{-1}}{m} = \frac{\bfD}{k_BT}
\label{covariance_dynamical}
\end{equation}
for particles of identical masses $m$. Thus, the eigenvalues $\{\Sigma_i\}$ of the position-position variance matrix and those of the dynamical matrix $\{\omega_i^2\}$ (the squared frequencies of the harmonic oscillators) are related by $\omega_i^2 = k_BT/(m\Sigma_i)$. However, this equality does not hold when assuming interatomic independence for $f_\textsf{GPP}$, as $\left[\boldsymbol{\Sigma}^{(\bfq,\bfq)}\right]^{-1}$ becomes a diagonal matrix composed of individual matrices $\Sigma_i^{(\bfq,\bfq)}$ (see \cref{GPP matrices}). 

\begin{figure}
\centering
\begin{tabular}{ccc}
\includegraphics[scale=0.5]{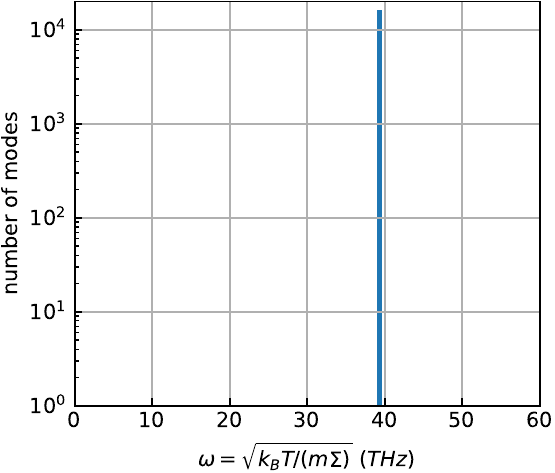} & 
\includegraphics[scale=0.5]{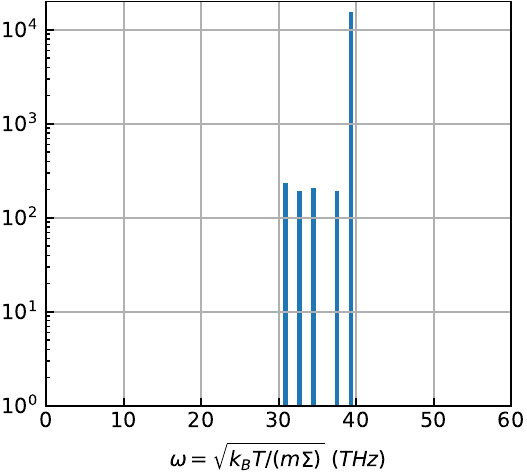} & 
\includegraphics[scale=0.5]{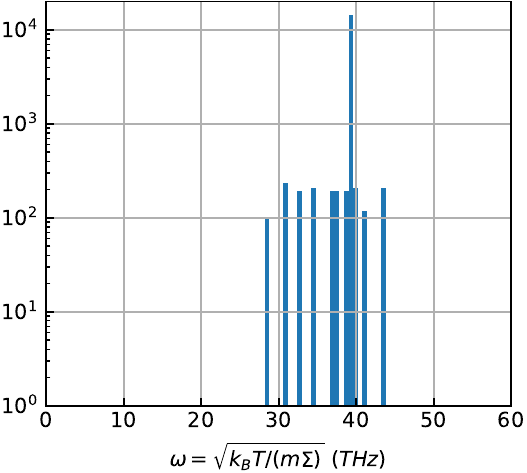} \\
\includegraphics[scale=0.5]{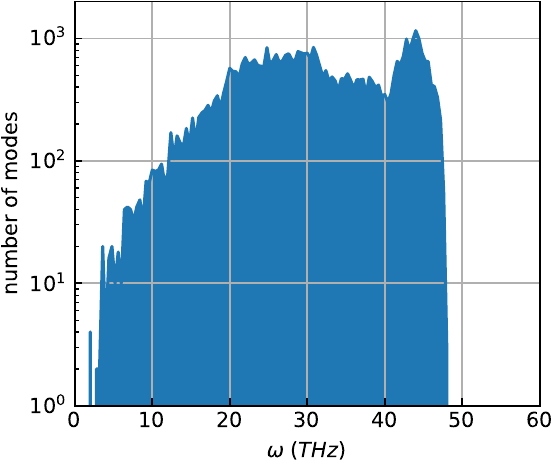} & 
\includegraphics[scale=0.5]{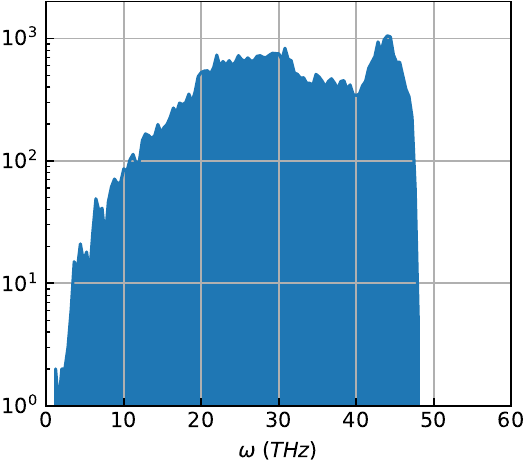} & 
\includegraphics[scale=0.5]{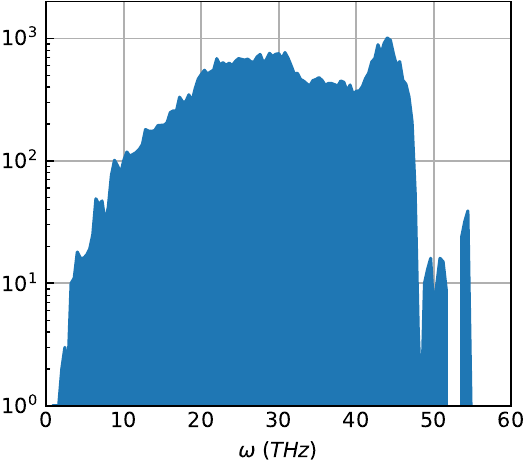} \\
(a) & (b) & (c)
\end{tabular}
\caption{Distributions of phonon mode frequencies for (a) a bulk sample, (b) a monocrystalline slab with two $\hkl[310]$ surfaces, and (c) a bicrystal with a $\Sigma5\hkl(310)\hkl[001]$ boundary and two $\hkl[310]$ surfaces, all obtained from the same atomic mean positions (obtained from relaxation with GPP) but using different approaches to compute the eigenfrequency spectra: (top tow) using GPP, (bottom row) using the QHA.}
\label{fig:dos_gpp_vs_qha}
\end{figure}

\begin{figure}
\centering
\begin{tabular}{cc}
\includegraphics[scale=0.65]{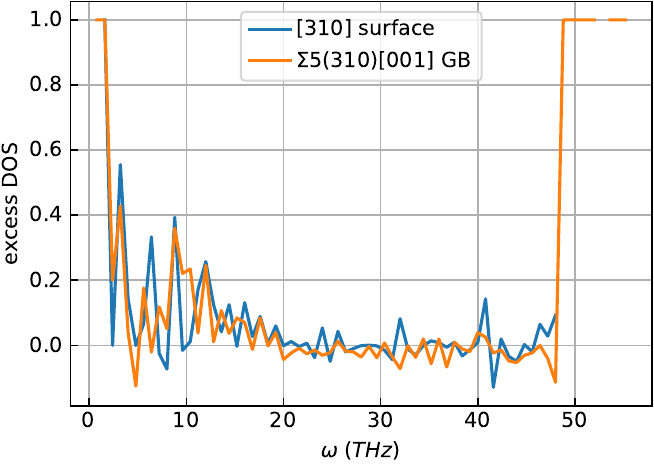 } & 
\includegraphics[scale=0.65]{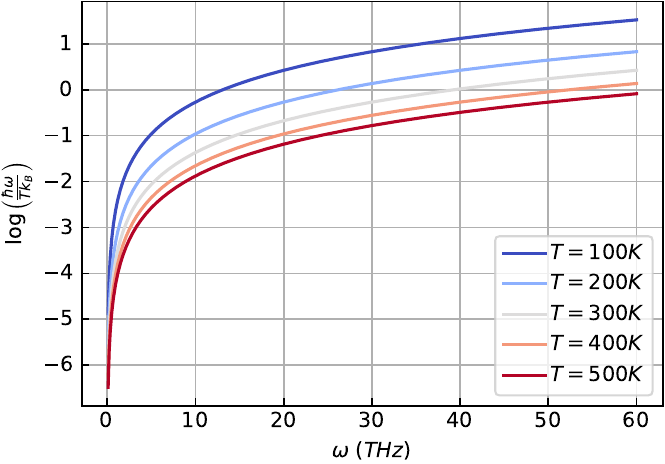}
\end{tabular}
\caption{Relative excess distribution of phonon mode frequencies for a $\hkl[310]$ surface and $\Sigma5\hkl(310)\hkl[001]$ boundary (left), and entropic contribution of each mode's frequency at different temperatures under the QHA (right). }
\label{fig:ex_vib_entropy}
\end{figure}

\begin{figure}
\centering
\includegraphics[scale=0.65]{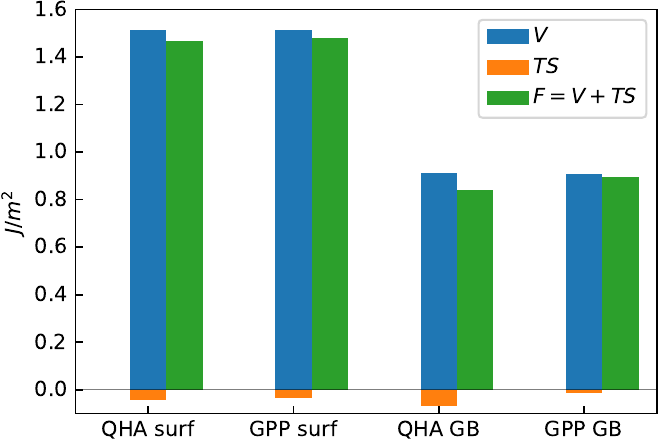}
\caption{Comparison of the different terms contributing to the free energy of a $\hkl[310]$ surface and $\Sigma5\hkl(310)\hkl[001]$ boundary obtained using the GPP framework and the QHA.}
\label{fig:compare_thermo}
\end{figure}

To shed light on the influence of this interatomic independence assumption on the vibrational spectra of atoms, we compare in Figure~\ref{fig:dos_gpp_vs_qha} two sets of vibrational frequencies. In both cases, structures are relaxed by GPP (assuming interatomic independence), so that the same mean atomic positions are used. The difference between GPP (top row) and QHA (bottom row) is the following. From GPP, the vibrational frequencies are obtained directly from the $\Sigma^{(\bfq,\bfq)}$ values resulting from relaxation, which assumes interatomic independence (violating the equality in \cref{covariance_dynamical}). From QHA, vibrational frequencies are obtained by computing $\bfD$ (\cref{dynmat}) by setting $\bfq_{i0},\bfq_{j0}$ to the mean positions obtained by the GPP relaxation and diagonalizing it. Both frequency spectrum calculations were applied to (a) a bulk sample, (b) a monocrystalline slab containing two $\hkl[310]$ surfaces, and (c) a bicrystal containing a $\Sigma5\hkl(310)\hkl[001]$ GB and two $\hkl[310]$ surfaces. As can be expected, the GPP output includes only a single frequency for the bulk, characteristic of an Einstein solid and resulting from the interatomic independence assumed in the position covariance matrix. Using the QHA for the bulk system results in the characteristic signature of the FCC density of states. For the monocrystalline slab, GPP captures the presence of lower frequencies, and for the bicrystal, it captures the presence of both lower and higher frequencies. While it is simple to recognize those additional frequencies introduced by the defects in the GPP data, is not as simple in the QHA vibrational spectra. As a remedy, we subtract DOSs in an analogous procedure as followed when subtracting free energies to extract surface and GB free energies in Eq.~\ref{eqF_gb_gpp}. The resulting excess DOS for each defect is presented in Figure~\ref{fig:ex_vib_entropy} together with the entropic contribution of each frequency. The data shows that the surface introduces lower frequencies compared to the bulk, while the GB introduces both lower and higher frequencies---which is consistent with the GPP observations. Compared to the monocrystalline slab, the excess DOS of the bicrystal involves more lower frequencies. Being a logarithmic function of the frequencies, the absolute entropy is more sensitive to changes in frequencies at the lower end of the spectrum, see Figure~\ref{fig:ex_vib_entropy}. Therefore, the free energy computed using the GPP-obtained frequency spectrum is more erroneous for a bicrystal (consisting of a GB) as compared to a monocrystalline slab with surfaces.   
Moreover, as shown by the QHA results, the GB excess entropy accounts for a larger fraction of the GB free energy than the surface excess entropy does for the surface free energy, so that inaccuracies of the entropy penalize the GB free energy more, as seen in Figure~\ref{fig:compare_thermo}. This proves that the interatomic independence assumed throughout this work introduces large errors in the excess vibrational entropies of GBs compared to those of free surfaces---even though relaxed atomic positions obtained from GPP are highly accurate. Therefore, we use the QHA to rebuild the vibrational spectra on GB structures that were efficiently relaxed by the GPP framework.

\section{Calculation of the GB free energy by thermodynamic integration} \label{appendixB}
We provide a brief description of the non-equilibirum thermodynamic integration (TI) method used to compute GB free energies by MD. This method was introduced in \cite{freitasNETI} and applied to compute the free energy of two different phases of a $\Sigma 5\hkl(310)\hkl[001]$ GB in \cite{freitasS5}. We adopt this method to compute the absolute free energy of different GBs by using three different samples: a bicrystal containing the studied GB and two free surfaces with equivalent crystallographic orientation as the GB, a monocrystalline slab containing the same free surfaces, and a bulk sample. The absolute free energies of these samples are used in \cref{eqF_gb_gpp} to extract the GB free energy. We repeat the TI calculation for each sample five times and take the average of the obtained absolute free energies as the final values used in \cref{eqF_gb_gpp}.

Thermodynamic integration is based on the Frenkel-Ladd method \cite{Frenkel19843188}: the change in free energy between two equilibrium states is computed by calculating the reversible work performed by taking a sample from an 'initial' state (for which we seek the absolute free energy) to a 'final' state (for which we know the free energy \textit{a priori}) through a series of equilibrium intermediate states forming the 'switching' path. The initial state is chosen to be described by a Hamiltonian $H_i$, assuming that all particles interact by the chosen interatomic potential, whereas the final state is described by an Einstein solid with Hamiltonian $H_f$, assuming that particles do not interact and behave as uncoupled harmonic oscillators with frequency $\omega$ about each lattice point. The final state hence has the a-priori known free energy
\begin{equation}
    \mathcal{F}_f(N, V, T) = 3Nk_BT\ln\left( \frac{\hbar\omega}{k_BT} \right).
\end{equation}
The switching path is parameterized by $H(\lambda)=(1-\lambda)H_i + \lambda H_f$, so that switching $\lambda$ from $0$ to $1$ takes the Hamiltonian from $H_i$ to $H_f$. At any point on the switching path, the free energy of the system is
\begin{equation}
    \mathcal{F}(N,V,T; \lambda) = -k_BT\ln \left(
    \frac{1}{h^{3N}} \int e^{-H(\lambda)/k_BT} \dd\bfp \dd\bfq
    \right).
\end{equation}It can be shown \cite{freitasNETI} that 
\begin{equation}
    \frac{\partial \mathcal{F}}{\partial \lambda} = \left\langle \frac{\partial H}{\partial \lambda}\right\rangle_\lambda = \langle H_i - H_f \rangle_\lambda,
\end{equation}
where $\langle \cdot \rangle_\lambda$ is a phase average under the canonical ensemble at fixed $\lambda$. Integrating this equation from $\lambda=0$ to $\lambda=1$ yields
\begin{equation}
    \mathcal{F}_i(N,V,T) = \mathcal{F}_f(N,V,T) + \int_0^1 \langle H_i - H_f \rangle_\lambda \text{d}\lambda,
\end{equation}
thus allowing to find the free energy of interest, $\mathcal{F}_i$, by tracking the phase average of the difference in Hamiltonians along the path. In equilibrium TI methods, the integral is numerically evaluated by discretizing the switching path in a series of equilibrium states with fixed $\lambda$-increments, each requiring a separate MD simulation to obtain $\langle H_i - H_f \rangle_\lambda$. As mentioned above, in this work we use non-equilibirum TI, which consists of a forward ($\lambda=0$ to $\lambda=1$) and a backward ($\lambda=1$ to $\lambda=0$) integration step, where the intermediate states for each $\lambda$ are not in equilibrium. This results in errors in $\langle H_i - H_f \rangle$ due to dissipation of the non-equilibrium nature of the process, which can be cancelled by combining the results of the forward and backward processes, if they are performed sufficiently slowly for linear response theory to be accurate. As a result, a single MD simulation is sufficient, which improves the efficiency. 

In practice, forward and backward switching were performed using the \texttt{ti/sping} fix \cite{freitasNETI} in LAMMPS \cite{LAMMPS}. Using this command, we equilibrated the states before forward and backward switching for $2\cdot 10^6$ timesteps, using a timestep size of $\dd t=1$~fs, and performed the switching in the same number of timesteps and with the same timestep size. The spring constant $k$ used to model the harmonic oscillators of the Einstein solid state was independently obtained from MD on a bulk sample in the canonical ensemble at each studied temperature by computing the mean-squared displacement (MSD) of the atoms and using the relation $k(T)=3k_BT/\textsf{MSD}(T)$ \cite{freitasS5}. Prior to switching, the selected initial geometries that yielded minimum-energy states using the GPP framework were relaxed under 0~K conditions. The dimensions of these initial geometries were expanded for the TI integration to achieve dimensions according to the convergence analysis of GB free energy performed by \citet{freitasS5} using TI for a $\Sigma5\hkl(310)\hkl[001]$. Thus, we used samples with dimensions $L_x,L_y,L_z$ larger than $72\si{\angstrom}$. Lastly, we note that the lattice spacing used for the samples in the TI simulations differs from the one used in the GPP relaxations, as explained in Section~\ref{conv analysis} (see also Table~\ref{tab:lattice_spacings}).

\end{document}